\documentclass[journal]{IEEEtran}
\usepackage{blindtext, graphicx}

\graphicspath{{./figures}}
\usepackage{booktabs}
\usepackage{epstopdf}
\usepackage{textcomp}
\usepackage{color}
\usepackage{enumerate}
\usepackage{mdframed}
\usepackage{footnote}
\usepackage{lipsum}
\usepackage{cite}
\usepackage{url}
\usepackage{tikz}
\usepackage{caption}
\usepackage{subcaption}
\usepackage{array}
\usepackage{multirow}
\usepackage{tabularx}
\usepackage[font=small,skip=0pt]{caption}
\usepackage{caption}
\usepackage{amsmath}
\ifCLASSINFOpdf
\else 
\fi

\begin{document}

\title{A Scalable Multi-Stage Packet-Switch for Data Center Networks}

	\author{Fadoua~Hassen,~\IEEEmembership{Student Member,~IEEE,} and Lotfi~Mhamdi,~\IEEEmembership{Member,~IEEE} 
		\thanks{The authors are with the School of Electronic and Electrical Engineering, Institute of Integrated Information Systems,University of Leeds, UK
			(e-mail:elfha@leeds.ac.uk; L.Mhamdi@leeds.ac.uk).}
	}
	
    \markboth{JOURNAL OF COMMUNICATIONS AND NETWORKS, VOL. X, NO. Y, DECEMBER 2016}{F. HASSEN and L. MHAMDI \lowercase{\textit{et al}}.: A Scalable Multi-Stage Packet-Switch for Data Center Networks} \maketitle

\maketitle


\begin{abstract}
The growing trends of data centers over last decades including social networking, cloud-based applications and storage technologies enabled many advances to take place in the networking area. Recent changes imply continuous demand for bandwidth to manage the large amount of packetized traffic. Cluster switches and routers make the switching fabric in a Data Center Network (DCN) environment and provide interconnectivity between elements of the same DC and inter DCs. To handle the constantly variable loads, switches need deliver outstanding throughput along with resiliency and scalability for DCN requirements. Conventional DCN switches adopt crossbars or/and blocks of memories mounted in a multistage fashion (commonly 2-Tiers or 3-Tiers). However, current multistage switches, with their space-memory variants, are either too complex to implement, have poor performance, or not cost effective.  We propose a novel and highly scalable multistage switch based on Networks-on-Chip (NoC) fabrics for DCNs. In particular, we describe a three-stage Clos packet-switch with a Round Robin packets dispatching scheme where each central stage module is based on a Unidirectional NoC (UDN), instead of the conventional single-hop crossbar. The design, referred to as Clos-UDN, overcomes shortcomings of traditional multistage architectures as it (i) Obviates the need for a complex and costly input modules, by means of few, yet simple, input FIFO queues. (ii)  Avoids the need for a complex and synchronized scheduling process over a high number of input-output modules and/or port pairs. (iii) Provides speedup, load balancing and path-diversity thanks to a dynamic dispatching scheme as well as the NoC based fabric nature. Simulations show that the Clos-UDN outperforms some common multistage switches under a range of input traffics, making it highly appealing for ultra-high capacity DC networks.

\end{abstract}


\IEEEpeerreviewmaketitle
\section{Introduction}

\begin{figure*}[tbh] 
	\centering
	\includegraphics[width=4.5in]{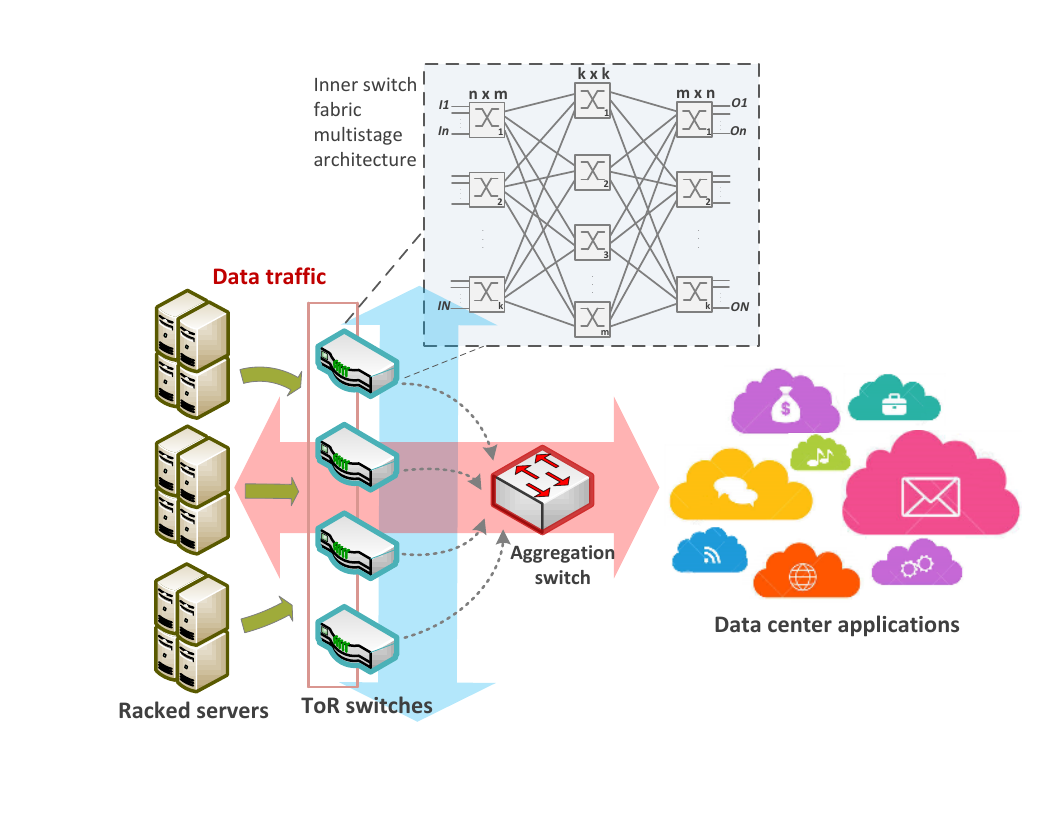}
	\caption{Abstraction of a ToR switching architecture in a Data Center Network } 
	\label{fig:data_center} 
\end{figure*}

\IEEEPARstart{I}{n} addition to virtualization, the on-going transformation of large-scale networks like DCNs is mainly realized through increasing the available bandwidth by means of high-performance and scalable switches/routers to ensure smooth and consistent enhancement for future needs. Targeting agility, Top of Rack (ToR) switching architecture can be adopted in a DCN for ease of network scaling and better power and cabling management (Fig.\ref{fig:data_center}). ToR switches need to be high-performance elements of high number of I/O ports to connect in rack servers. In this context, aggregate throughput, latency, complexity and power consumption are key considerations when designing switches for DCN environment\cite{art51}. Commonly, hierarchical switching fabrics are built to manage the floating traffic in DCNs. Single-stage crossbar switches do not meet the growing networking requirements. While they can be implemented for small-sized switches, they become quite complex to implement and unscalable for growing port counts (beyond 64 ports)\cite{art1}\cite{art17}. Other design approaches have been investigated, such as multistage switches where many smaller crossbar fabrics are arranged in cascade. They have been typical commercial solutions for high-speed routers \cite{art18} mainly because they can be incrementally expanded by adding more modules to the existing design. Besides, they have numerous benefits such as being partially or completely non-blocking, providing good broadcast and multicast features and build in reliability with no or minimum failure in the system. Dell S-Series 100M/1G/10G/40GbE ToR switches have been designed and optimized to leverage a non-blocking architecture that delivers low-latency switching and increase scalability at the DC network edge \cite{art50}. Cisco designed the Nexus 5000 Series ToR switches to support flexible deployment and to meet the scalability demands of today's data centers \cite{art51}. Juniper also provides the EX4550 line switches that fit for high density data center ToR deployments \cite{art52}.

One of the most popular multistage arrangements is the three-stage Clos-network that is frequently used for telecommunications and networking systems \cite{art2} \cite{art8}. The buffer placement defines the type of the multistage Clos switch which can be a Space-Space-Space (S$^3$) \cite{art16} network without buffers or Memory-Memory-Memory (MMM) \cite{art7} with buffered switching units in all stages. Other combinations have also been studied \cite{art3} \cite{art6}. Despite their scalability potential, almost all existing Clos-network based proposals are either too complex to be implemented, exhibit prohibitively high cost or have poor performance. The input queuing structure at the input modules (IMs) is generally complex, requiring excessive number of queues to avoid the Head-of-Line (HoL) blocking \cite{art3}\cite{art7}.
In addition to their impact on the scheduling complexity, these queues are generally required to be of output queued type and run much faster than the external input line rate. On the other hand, the scheduling process in current multistage Clos-networks, especially the  Memory-Space-Memory (MSM) type, is very complex and expensive, yet has poor performance under non-uniform traffic scenarios. A typical example of this is the Concurrent Round Robin Dispatching (CRRD) for MSM and its enhanced versions \cite{art4} \cite{art5}. MMM packet switches involve many buffers at all stages of the Clos-network \cite{art7} to relax the scheduling complexity which lead to prohibitively increasing the implementation cost.

Irrespective of whether the switch is single or multistage, it is constructed of either one or multiple single-hop crossbar fabrics as its elementary building block. In this context, the Network-on-Chip (NoC) paradigm has recently been gaining interest in  modern high-performance single-stage switching fabrics design as it addresses a number of limitations of conventional single-hop crossbars, including scalability, port speed and path diversity \cite{art13}.  A number of recent designs have used the NoC concept in high-performance switching. A design for Ethernet switches has been described in \cite{art23}\cite{art24}.  A Unidirectional NoC crossbar fabric based packet switch (UDN) design has been described in \cite{art13} \cite{art15} along with appropriate NoC routing algorithms. An extension of this design, termed MultiDirectional NoC (MDN) packet switch has been also proposed in \cite{art14}. More recent results \cite{art22} proposed an implementation of a single-stage crossbar fabric using NoC-enhanced FPGA and different routing algorithms. Despite the high potential of NoC based crossbar fabrics, their application has been restricted to single-stage crossbar packet switches. 

In our previous work, we proposed the first design of a scalable multistage packet-switches based on NoC fabrics for DCNs \cite{art36}. In particular, the switching fabric is a combination of a Clos macro-design, that reports to the whole fabric architecture, and a UDN micro-design for the central switching modules of the packet-switch. We describe a three-stage Clos-network with FIFO input queues and a dynamic dispatching of packets to the central modules. The proposed switching architecture has several advantages over earlier multistage packet-switches. In particular:
\begin{itemize}\renewcommand{\labelitemi}{$\bullet$}
\item{
	The Clos-UDN obviates the need for a complex and costly input queuing structure. Unlike conventional multistage design where a high number of fast Virtual Output Queues (VOQs) is required, the Clos-UDN uses a small number of input FIFO queues which need not to run faster than the external line rate. 
	}
\item{
	The proposed Clos-UDN avoids the need for complex, costly and slow centralized scheduling process. Conventional multistage Clos-networks require complex scheduling process with global synchronization between inputs and outputs. 
	Our proposal relies on the UDN stages to route input packets to their outgoing interfaces by means of fully distributed, parallel and independent NoC routers' decisions.
	}
\item
	{The Clos-UDN inherits all the advantages of the UDN design in terms of scalability, speedup and path diversity\cite{art13}. These properties result in high performance in terms of low latency, high-throughput and efficient hardware design.
	}
\end{itemize}

Out-of-sequence packets delivery is a common problem to all multistage packet switch architectures with buffered middle stages.  A re-sequencing mechanism at the output stage of the switch \cite{art7} is a popular solution to this phenomenon. In\cite{art7}, are discussed two re-ordering mechanisms based on time-stamp monitoring that is performed either at the input modules (MMM-IM switch) or at the output modules (MMM-OM switch). Although both alternatives do not require any synchronization among the different blocks, many buffers and arbiters have been introduced making the solutions unscalable. In\cite{art8}, H. J. Chao $et~al.$  proposed a multi-plane, multistage buffered switch with several re-sequencing mechanisms including: Static and dynamic hashing, time stamping and window-based re-sequencing. It is worth mentioning that almost all previously suggested approaches rely on complex algorithms \cite{art32} and imply the use of numerous schedulers and buffers\cite{art7}\cite{art34}. We suggest a simple way to alleviate the packets mis-sequencing in the Clos-UDN switch. We show that a static configuration of the input and the central stage modules connections used along with the appropriate routing algorithm across the UDNs grantees an ordered packets transfer. 

The reminder of the paper is structured as follows. Section \ref{Sec.RW} discusses relevant existing multistage Clos packet-switch architectures and their performance. In Section \ref{Sec.Clos-UDN}, we describe the three-stage Clos-UDN packet-switch architecture, along with its NoC based central modules and its dispatching process. Section \ref{Sec.Hardware_cost} overviews some hardware requirements of the proposed switch and compares them to those of MSM and MMM switches. In section \ref{Sec.out-of-seq} we present the static packets dispatching scheme that ensures an ordered packets delivery. Section \ref{Sec.Sim} is reserved for the performance study of the Clos-UDN switch and section \ref{Sec.Conc} concludes the paper.

\section{RELATED WORK} \label{Sec.RW}
Multistage network switches are more scalable than single stage crossbars. They are used in large-scale networks like DCNs for their scalability and reliability. They provide multiple routes between inputs and outputs, allowing the traffic to be balanced across alternative paths. Non-blocking Clos-network is a very popular design \cite{art2}. A three-stage Clos is generally quoted as  $\zeta(\it{m},\it{n},\it{k})$   where $\it{m}$, $\it{n}$, and $\it{k}$ are the parameters that completely define the structure of the network. The size of this Clos-network is $N$, where $N = (n\times k$). The first stage is made of  $\it{k}$ input modules each of size ($n\times m$). The middle stage has  $\it{m}$ switches each has  $\it{k}$ inputs and  $\it{k}$ outputs. Last, there are  $\it{k}$ output modules at the third stage, each of size ($m \times n$). Extensive work has been done on Clos-network switches in all their variants such as S$^3$ \cite{art16}, MSM \cite{art3}\cite{art4}\cite{art5}, Space-Memory-Memory (SMM) \cite{art6} and MMM \cite{art17} \cite{art7}. 
Unfortunately,  none of the existing Clos-network switching architectures has been shown to provide scalability in terms of cost, performance and hardware complexity. The MSM architecture requires expensive and complex input modules. Each of these input modules is required to cater for a high number of separate FIFO queues ($n.k$) in order to avoid the HoL blocking. Additionally, each of these queues is required to run ($n+1$) times the line rate \cite{art4}. On the scheduling/dispatching front, the cost and practicality is a major issue. Two scheduling phases are required to resolve the input-output ports contention. In addition to its high cost and long scheduling delays, no scheduling algorithm for this architecture has been shown to exhibit satisfactory performance \cite{art3} \cite{art5}. As for MSM, MMM has $N$  VOQs at the IMs to prevent HoL blocking.
The buffered architecture \cite{art7} \cite{art17} mandate expensive internal memories to relax the scheduling process. Fully-buffered Clos architectures have good throughput performance since all contentions are absorbed by means of internal buffers. Although the scheduling process is better than that of MSM, it is still complex.  
\begin{figure*}[tbh] 
	\centering 
	\includegraphics[width=5.5in]{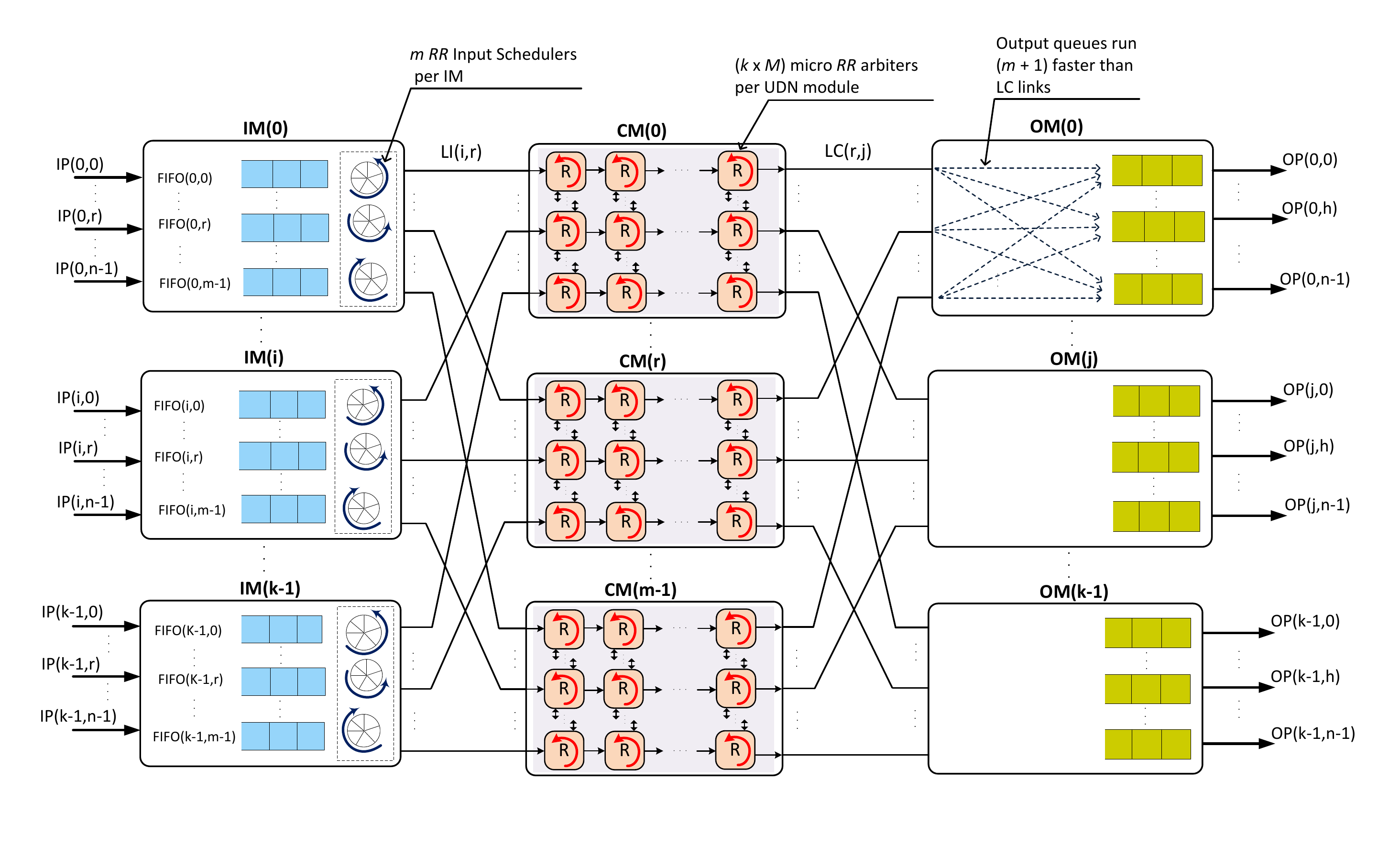}
	\caption{($N\times N$) three-stage Clos-UDN packet-switch architecture with dynamic dispatching scheme} 
	\label{fig:clos_switch} \vspace{-.3cm}
\end{figure*}

Our work differs from all previously proposed architectures. We take a radically different approach at the heart of Clos-switch design by adopting NoC based fabrics as internal stages of the Clos-network. Designing each Central Module (CM) as a NoC brings a number of advantages that overcome the limitations of previous proposals. First, the input modules are less complex and cheaper compared to previous architectures. Each input module of the  Clos-UDN switch requires only $m$ input FIFO queues, each of which runs \textit{twice} the line rate. This is to be compared to the MSM and MMM, where each input module requires ($n.k$) input FIFO queues each of which runs ($n+1$) times the line rate. Contrary to the complex, costly and under-performing proposed schedulers in traditional Clos architectures, the Clos-UDN uses fully distributed and parallel scheduling at the NoC routers level, making it simple, fast and efficient as we shall describe next. 

The dynamic cells dispatching scheme disorders packets by distributing them to different paths across multiple UDNs through time. By imposing a static configuration of the input and central modules links, the number of stages of the Clos-network gets reduced to two instead of three. Two-stage interconnects are more scalable than single stage architectures. However they are blocking if only one link is used to connect any input/output pairs of modules. Hence, connections must be built in redundancy and a large number of links between the first and second modules is required. The load balanced Birkhooff-Von Neumann switch made with two stages of crossbar switches was proposed in \cite{art31}. The switch is made with input-buffered modules in the second stage. It has no schedulers and adopts a deterministic sequence of $N$ different configurations to connect $N$ input/output pairs of modules. A disadvantage of the two-stage switch is that it can experience ou-of-sequence packets delivery \cite{art32}. To prevent packets mis-sequencing and to maintain performance benefits of the two-stage switch, I. Keslassy $et~al.$ introduced expensive three dimensional queues (3DQs) and a frame-based scheduling algorithm \cite{art32}. In a different approach to build scalable switches, R. Rojas-Cessa  $et~al.$ discussed a bufferless two-stage scalable switch with module-first matching scheme \cite{art33} where an iterative matching is performed between the input and output modules at the first place and ports matching occurs later. Although they are interesting, two-stage interconnects have several limitations which urged other architectures to rise. In the current work, we impose a  static configuration for the IMs and CMs connections. Subsequently, packets of the same flow stored in the same input FIFO are constantly sent to the same UDN block where they are routed to their corresponding outputs in the order of their arrivals. We show in section \ref{Sec.out-of-seq} shows that a static configuration simplifies the switch and preserves good performance under a range of traffic types while preserving packets order. Additionally, input schedulers are no more needed and the switch architecture can be viewed as a two-stage interconnect with in-order packets delivery guarantee.     

\section{Clos-UDN SWITCH ARCHITECTURE} \label{Sec.Clos-UDN}
This section describes the three-stage Clos-UDN switch architecture with NoC-based central modules. We describe the switch model with an emphasis on the NoC based central modules. We then introduce the dispatching process considered to transfer packets to the middle stage.
\subsection{The switch model}
The reference design of a Clos-UDN switch of size ($N\times N$) is depicted in Fig.\ref{fig:clos_switch}. The key notations used in this paper are listed as in TABLE \ref{tab:term}. The first stage of the Clos-UDN comprises $\it{k}$ IMs, each of which is of size ($n\times m$). The second stage is made of $\it{m}$ UDN fabric modules, each of dimension\footnote{Unlike conventional Clos-networks, the central modules of the Clos-UDN can be of size ($k\times M$) crosspoints, where $M$ refers to the NoC depth and $M \leq k$.} ($k\times k$). The third stage consists of $\it{k}$ OMs, each of which has ($m\times n$) dimension.
Although it can be general\footnote{The Clos-UDN can of course be of any size, where $m \geq n$. This would simply require packets insertion policy in the FIFOs should we need to maintain low-bandwidth FIFOs. We consider this to be out of the scope of the current work.}, the proposed Clos-UDN architecture  has an expansion factor $\frac{m}{n} =1$, making it a $\it{Benes}$ lowest-cost practical non-blocking fabric. An IM($\it{i}$) has $\it{m}$ FIFOs each of which is associated to one of the $\it{m}$ output links denoted as LI($\it{i}$, $\it{r}$).  An LI($\it{i}$, $\it{r}$) is related to an CM($\it{r}$). Because $m=n$, each FIFO($i,r$) of an input module, IM($\it{i}$), is associated to one input port, IP($i, h$), and can receive at most one packet and send at most one packet to one central module at every time slot.
\begin{figure}[tbh]
	\begin{center}
		\includegraphics[width=3.7in]{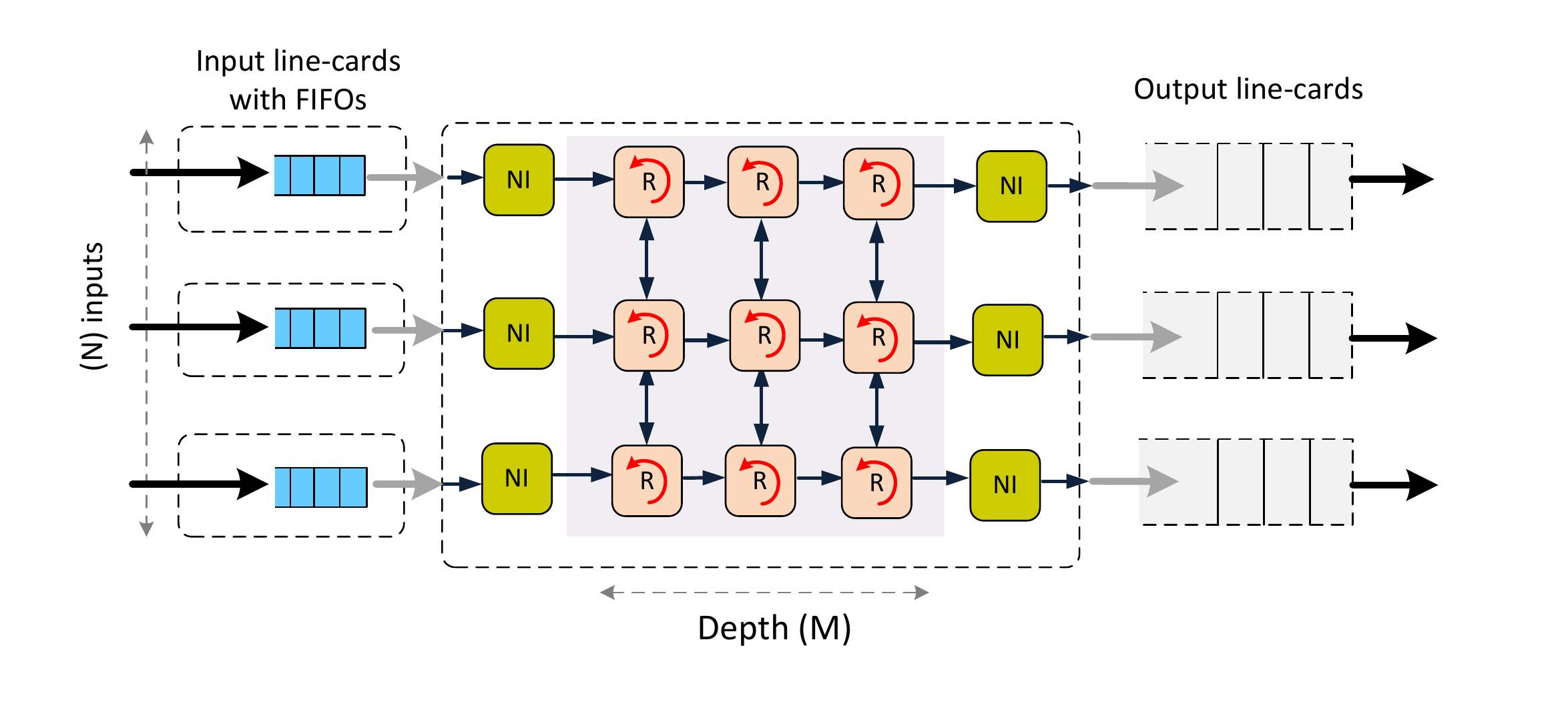}
	\end{center}
	\caption{The UDN crossbar switch}
	\label{fig:noc}
\end{figure}
A CM($\it{r}$) has $\it{k}$ output links, each of which is denoted as LC($\it{r}$, $\it{j}$) and is connected to OM($\it{j}$). An OM($\it{j}$) has $\it{n}$ OPs, each of which is OP($\it{j}$, $\it{h}$) and has an output buffer. An output buffer can receive at most $\it{m}$ packets and forward one packet to the output line at every time slot. Packets destined to different output ports are accepted to the NoC fabric even when some outputs are busy with other packets. 

\vspace{.2cm}
{\small	
	\centering
	\captionof{table}{The terminologies for the Clos-UDN switch} \label{tab:term} \vspace{.3cm}
\setlength{\tabcolsep}{1em} 
{\renewcommand{\arraystretch}{1.2}

\begin{tabular}{p{1.6cm}p{6cm}} 
	\hline
	\specialrule{.05em}{.05em}{.05em} 
	Notation    & Description \\
	\hline
	\specialrule{.05em}{.05em}{.05em} 
	IM($i$)	&  ($\it{i}+1$)$^{th}$ IM at the first stage \\
	CM($r$)	&  ($\it{r}+1$)$^{th}$ CM at the second stage \\
	OM($j$)	&  ($\it{j}+1$)$^{th}$ OM at the third stage \\
	\it{i} &  IM number, where  ${0 \leq i \leq k-1}$\\
	\it{r} &  CM number, where  ${0 \leq r \leq m-1}$\\
	\it{j} &  OM number, where  ${0 \leq j \leq k-1}$\\
	\it{h} &  IP/OP number in each IM/OM, respectively, where ${0 \leq h \leq n-1}$\\ 
	IP($i, h$)	&  ($\it{h}+1$)$^{th}$ IP at IM($i$)\\
	OP($j, h$)	&  ($\it{h}+1$)$^{th}$ OP at OM($j$)\\
	FIFO($i, r$) &  First-In-First-Out queue that stores packets going to CM module, $r$.\\
	LI($\it{i}$, $\it{r}$)	&  Output link at IM($\it{i}$) that is connected to CM($\it{r}$)\\
	LC($\it{r}$, $\it{j}$)	&  Output link at CM($\it{r}$) that is connected to OM($\it{j}$)\\
	\hline
	\specialrule{.05em}{.05em}{.05em} 	
\end{tabular}
} 
\newline
\\
}

\subsection{NoC based Central Modules}
Our reference design is based on the UDN \cite{art13} fabric (Fig. ~\ref{fig:noc}) that we plug into the Clos central stages. In the Clos-UDN, every central unit is a two-dimensional mesh ($k\times k$) of small on-chip packet switched input-queued routers that transport packets across the NoC in a multi-hop fashion. All on-chip routers have small input FIFO queues of variable size (referred to as Buffer Depth- $\it{BD}$) to store packets on their journey to their outputs. To avoid elastic buffers, credit-based flow control is used and packets are only sent when buffer space is available \cite{art12}. A packet is of fixed-size with relative routing information stored at its header. Packets are fully received and stored in one of the router's buffers before going to the next hop. Using a deadlock-free NoC routing algorithm, named $"Modulo~XY"$ \cite{art14}, packets advance in the NoC fabric at a rate of one packet per time-slot \cite{art14}. We define the speedup of a UDN module as the speed ratio at which the fabric can run with respect to the input/output ports. It is equivalent to the fabric removing up to $\it{SP}$ packets from one input and sending up to $\it{SP}$ packets to one output per time slot. Unlike the centralized decision making in other multistage packet switch architectures, on-chip routers make local decisions about the packets next destinations using RR arbitration making the scheduling process distributed. The UDN switch can sustain high throughput and low delays under heavy loads if the fabric is running with a small speedup ($\it{SP}{>}$1). Given the small sized on-chip routers and short wires, a speedup of 2 can be readily affordable. Sections~\ref{Sec.Sim} and \ref{Sec.Hardware_cost} further study this property.

\begin{figure*}[tbh] 
	\centering
	\includegraphics[width=5.5in]{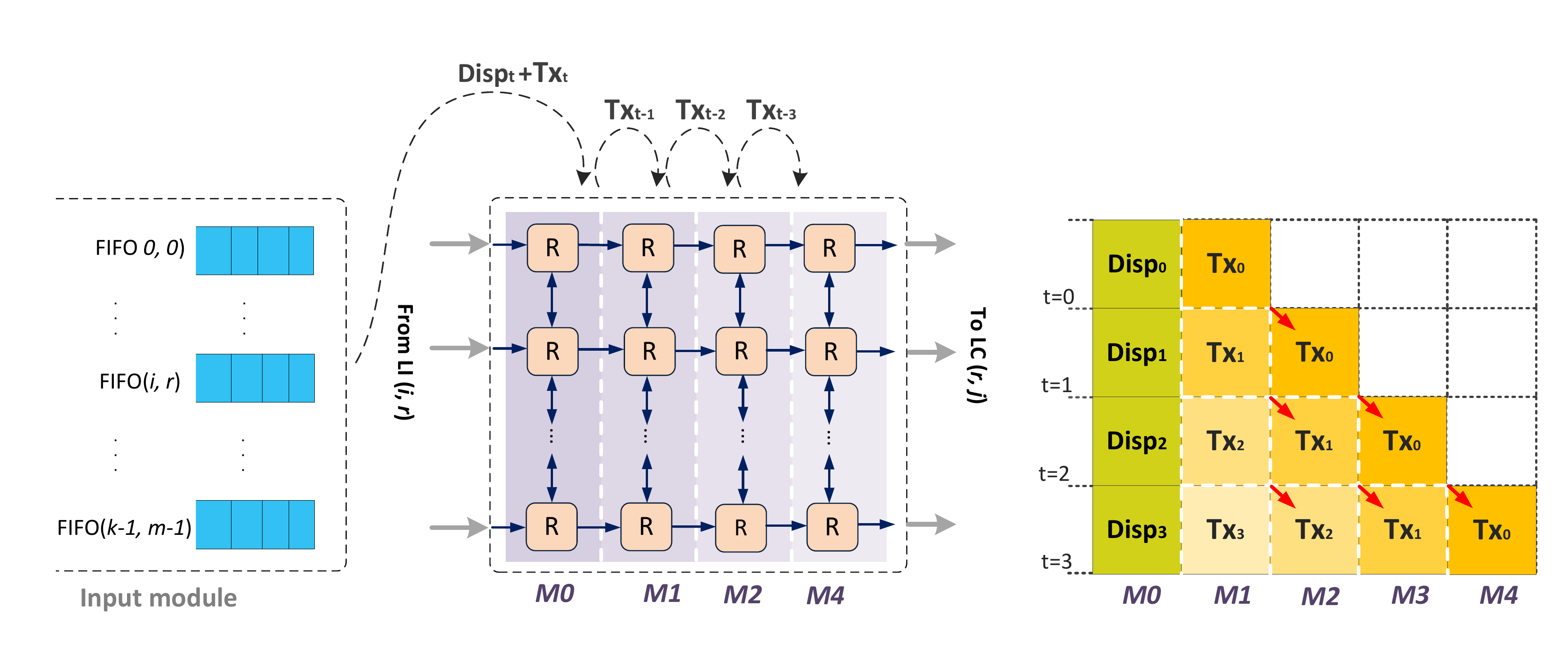}
	\caption{Pipelined working of Clos-UDN dispatching and packets forwarding through UDN modules.} 
	\label{fig:pipelining} \vspace{.3cm}
\end{figure*}

\subsection{Head of Line Blocking}

Common single-hop crossbars experience HoL blocking whenever packets wait in the line cards for their corresponding output ports to be available. These packets block other cells that are queued behind them in the same line card, even though the latter are destined for free output ports\cite{art13}. Multi-hop NoC-based crossbars such as the UDN modules, do not suffer the HoL limitation because of the multistage and pipelined nature of the NoC itself. Packets from a single input port heading to different output ports can be accepted into the NoC structure even if their outputs are busy. Moreover, the geometric features of the NoC-based structures (path diversity) and adequate routing methods contribute towards better load balancing. The traffic is parallelized on multiple paths and packets which intend to go to different output ports interfere little with each other. Thus, one can use FIFOs instead of VOQs on the line cards with no severe performance degradation\cite{art53}.

\subsection{Packets scheduling / dispatching in multistage switches}
The need for a conflict-free matching in conventional Clos-network switches, such as MSM, mandates the need of two types of matchings \cite{art3} \cite{art4} \cite{art5}: a matching within each input module to select eligible VOQs among non-empty candidate VOQs and a second matching between IMs and CMs. Both of these matchings are quite complex and time consuming due to the high number of input queues per IM for the first matching as well as the global synchronization of input-output port pairs for IM-CM matching to produce a conflict-free match. Buffers at all stages of an MMM switch absorb the contention and obviates the need for a complex and centralized matching between modules and links of the Clos-network. Larger internal buffers make MMM perform better especially under heavy input loads and bursty traffic. However, modern integrated circuit technology still limits concrete buffered fabrics production. Some proposals suggest pipelined schedulers to manage packets journey in a structure with $b$ sized crosspoint buffers using rigorous, yet complex scheduling scheme\cite{art17}.  

The proposed Clos-UDN greatly simplifies the process of packets dispatching and scheduling. First, each IM needs to maintain only $m$ input queues and each input port of an input module can send to only one FIFO queue per time-slot, making the FIFO running at only twice the line rate. In the Clos-UDN switch, there are $m$ input schedulers in every IM, one per FIFO queue. The RR input schedulers are initialized to different values and they keep updating their selection pointers to one position at the end of every time slot. This guarantees that all pointers are always desynchronized and no conflict in the LI links happens. At the start of every time-slot, a scheduler selects an LI($\it{i}$, $\it{r}$)  link among $\it{m}$ links in a RR fashion to transfer the HoL packet from each non-empty FIFO to a central stage/module of the Clos-UDN network. A packet is accepted to the CM module if the left-most NoC router still has room in its left buffer. Once at the NoC, the  $"Modulo~XY"$ routing algorithm takes over and routes the packet to its outgoing port.
MMM as described in \cite{art17} employs a single central admission scheduler that manages all credit and grant sub-schedulers operating in a pipelined way. Although the suggested scheduler has softer operating time than in MSM architecture, it is more complex than the scheduling scheme the Clos-UDN switch.  Schedulers in the line cards send requests to central arbiter. Credit schedulers associated to the switch outputs manage as many credit counters as crosspoints and allocate buffer space for packets. Ultimately, grant schedulers select one among candidate requests and send back a grant signal to the central admission scheduler.
Our proposal differs from other types of Clos-network switches with bufferless and buffered middle stages. Mainly, our architecture does not require an IM-CM matching. The central modules, NoC fabrics, make parallel and distributed forwarding decisions independently.  Routers of the UDN decide about the next hop of transferred packets. They examine and modify the route information continuously until the packet reaches its destination. Packets contending for a link would remain stored to the router's buffers before the arbiter grants them access \cite{art13}\cite{art15}. Correspondingly, contention for LC($\it{r}$, $\it{j}$) links gets resolved within the UDN units as packets progress in the NoC as Fig.~\ref{fig:pipelining} shows. Hence, the process of path-allocation in the Clos-network is relaxed and no centralized and global decision and synchronization are needed. 

\section{Hardware requirements}\label{Sec.Hardware_cost}
In this section, we briefly compare the hardware requirements of the proposed Clos-UDN switch to MSM and MMM switches. CRRD scheduling scheme and its derivatives as adopted for MSM, perform iterative matchings between the set of eligible VOQs and the available LI links. In MMM architecture, packets are selected in a RR manner to move to the available crosspoint buffers before they undergo another selection to be transferred to the output buffers in the OMs. TABLE \ref{table:1} compares some features of the three switching architectures.

	\begin{table*}[!htbp]
		\centering		
		\setlength{\tabcolsep}{.5em} 
		{\renewcommand{\arraystretch}{1.3}
			\caption{Compare the HW requirements of MSM, MMM and Clos-UDN switches}	\label{table:1}
			\captionsetup{belowskip=4pt,aboveskip=5pt}
			\begin{tabular*}{1.5\columnwidth}{c| c| c| c| c} \cline{2-4}
				& MSM  & Three-stage Clos-UDN & MMM & \\ \cline{1-4}
				\multicolumn{1}{|c|}{Input buffers per IM} & ($n \cdot k$) VOQs & $m$ FIFOs &  ($n \cdot k$) VOQs + ($n \cdot m$) VCMQs& \\ \cline{1-4}
				\multicolumn{1}{|c|}{Central modules} & Bufferless  &  NoC-based UDN    & buffered   & \\ \cline{1-4}
				\multicolumn{1}{|c|}{Dispatching scheme} & CRRD  &   Dynamic RR dispatching  & RR / LQF  & \\ \cline{1-4}
				\multicolumn{1}{|c|}{Contention resolution} & Two phase matching  &   NA   & NA  & \\ \cline{1-4}
				\multicolumn{1}{|c|}{IM arbiters} & $N + m$ &    $m$ & $n + m$ & \\ \cline{1-4}
				\multicolumn{1}{|c|}{CM arbiters} & $m$ &   NA & $k$ &  \\ \cline{1-4}
				\multicolumn{1}{|c|}{Scheduling algorithm} & Centralized/complex &   Parallel / distributed & Distributed & \\ \cline{1-4}
				\multicolumn{1}{|c|}{In-order packets delivery} & Yes &    No & No & \\ \cline{1-4}
				\multicolumn{1}{|c|}{Parameterization} & NA   &   YES   &  YES  & \\ \cline{1-4}
				\multicolumn{1}{|c|}{Performance / scalability  } & low   &    High  & Good  & \\ \cline{1-4} 
			\end{tabular*}	
		}	
	\end{table*}

We mention that a prototype of the UDN crossbar switch has been synthesized in our previous works \cite{art15} using an ASIC 65 nm CMOS technology. The synthesis of a ($3 \times 3$) UDN switch with no optimization measures achieved 413 MHz with a cell area of 4.8 mm${^2}$ including the Network Interfaces (NIs). As defined in the same reference, the degree of a router is the number of its I/O ports. The UDN fabric has 3 degree (in the east and west mesh columns) and 4 degree (in the intermediate mesh columns) NoC routers that occupy respectively 0.29 mm${^2}$ and 0.38 mm${^2}$. The NIs occupy 0.32 mm${^2}$ considering the same synthesis technology \cite{art13}. Namely, registers that are used for the routers' FIFO queues dominate the area. The die area of the circuit is shown to drastically shrink if dedicated hardware rippled-through FIFOs and other CMOS process technologies are used (e.g. The area of an $N=M=32$ switch using CMOS 65 nm process is 403 mm${^2}$ and only 134 mm${^2}$ if 90 nm CMOS technology is used). Adopting 65 nm CMOS process, a central module of size ($8 \times 8$) used in a ($64 \times 64$) Clos-UDN network switch would occupy: ${0.29\times2M + 0.38(N-2)M + 0.32\times2N \approx 224}$ mm${^2}$/CM.

\subsection{Dispatching time}
\subsubsection{MSM with CRRD dispatching}
CRRD has two-phase matching process: Matching within IMs and the IM-CM matching. Phase 1 is an iterative matching that runs $iter$ times to maximize the subset of connected VOQs to the output-links LI(\textit{i} , \textit{r}). At every iteration, two RR arbiters are used as follows: An output link LI(\textit{i} , \textit{r}) selects one out of at most ($n.k$) requesting VOQs. A VOQ arbiter chooses one among at most \textit{m} grants. The resulting complexity in phase 1 is $\mathcal{O}(iter~(log~nk))$ where $iter$ is the number of iterations ($1\leq~iter\leq~m$). During the IM-CM matching, every LC(\textit{r}, \textit{j}) arbiter chooses one among at most \textit{k} requests. The time complexity of this phase is then $\mathcal{O}(log~k)$.

\vspace{.3cm}
\subsubsection{Packets dispatching in MMM }
Arriving packets get stored in VOQs at the input ports of the MMM\footnote{MMM packet switch architecture as described in \cite{art26}.} switch. There is a total of $N$ arbiters, one per input port, to select the cell to send to one of the  $m$ Virtual CM queues, VCMQs, at the IMs \cite{art26}. The selection of VCMQs is RR based. A total of $m$ arbiters in each IM are used to perform the selection of the CM through which the cell is sent. We conclude that the dispatching complexity is  $\mathcal{O}(log~nm)$\footnote{As we are considering only non-blocking Clos-networks, $m \geq k$.}. 
\vspace{.3cm}
\subsubsection{Dispatching in the Clos-UDN switch}
The dispatching scheme is non-iterative and made of a single phase. At each cell time, $m$ RR arbiters in the IMs select CMs to dispatch HoL packets. This makes the complexity time equal to $\mathcal{O}(log~m)$. The dispatching process and packets routing through the UDN modules work in parallel. The pipelined nature of the UDNs makes the dispatching time at time slot $t$ ($Disp_t$) and the packets scheduling and forwarding through the NoC ($Tx_t$) overlapping. We call $F_0$, the flow of packets dispatched to a particular UDN module at time slot $t=0$. As depicted in Fig.~\ref{fig:pipelining}, $F_0$ arrives to the NoC routers of the first column $M_{0}$. Forwarding decisions are taken and packets are transferred to inputs of the next hop. At time slot $t=1$, a new flow of packets $F_1$ arrives to $M_{0}$ while $F_0$ gets routed to the next stage of the UDN. Solidly connecting the two first stages of the Clos-UDN switch removes the ($k \times m$) input schedulers used to dynamically dispatch cells to the central modules. Henceforth, the complexity of the switch is considerably reduced.

\subsection{Hardware complexity for dispatching }
\subsubsection{MSM with CRRD dispatching scheme}
Every IM has \textit{m} output-link arbiters and \textit{N} VOQ arbiters. Generally, the complexity of the RR arbiter is $\mathcal{O}(log~n_{req})$,
where \textit{n}$_{req}$ is the number of requests to be selected by the arbiter. The hardware complexity in case of MSM with CRRD scheme in the IM is $\mathcal{O}(log ~mN)$\cite{art4}. There are also \textit{k} LC(\textit{r}, \textit{j}) arbiters at the central modules each of which selects one request out of at most \textit{k} requests. The total complexity of a CM is $\mathcal{O}(log~k)$.
\vspace{.3cm}
\subsubsection{MMM switch}
MMM architecture requires many buffers to house packets at every stage of the Clos switch. In addition to the $N$ VOQs present at the input ports, each IM block is made of ($n\cdot m$) VCMQs. The operations of VOQ and VCMQ selection performed before the packets dispatching to the CMs, make the hardware complexity of an input module of MMM switch $\mathcal{O}(log~nN)$. The  HW complexity of a CM and OM are $\mathcal{O}(log~k^2)$ and $\mathcal{O}(log~nm)$ respectively.
\vspace{.3cm}
\subsubsection{Clos-UDN switch}
In Clos-UDN switch, \textit{m} arbiters per IM are associated to \textit{m} FIFOs. 
A queue arbiter selects one among \textit{m} CMs to dispatch the current HoL packet which makes the hardware complexity of an IM equal to 
$\mathcal{O}~(log~m)$). Every CM bloc at the central stage is made of ($k\times M$) mini-routers. Every on-chip router selects packets in a RR manner to forward them to the next hop. This results into a complexity of $\mathcal{O}(log~kM)$. 
%

During the first phase of the CRRD matching, requests are sent to the LI(\textit{i}, \textit{r}) arbiters that send back grants to the selected VOQs. A VOQ arbiter accepts one among the received grants. The size of the interconnect between \textit{m} output-links and ($n\cdot k$) VOQs arbiters increase with the switch size making the wiring more complex. The number of crosspoints $N_{xp}$ for interconnection wires between the IM arbiters is given by $N_{xp}= \frac{3}{4} nkm(nk -1)(m-1)$ \cite{art4}. To diminish the layout complexity, the CRRD dispatching scheduler needs to be done on multiple-chips\cite{art4}. Still, the interconnection between the chips on Printed Circuit Boards (PBCs) and the number of pins in the scheduler chips becomes higher and expensive.
Clos-UDN has a simple dispatching scheme and the IM arbiters are not connected to any others which saves the need for complex interconnects. Such a feature makes the Clos-UDN scalable independently of the switch size. The contention resolution in Clos-UDN is progressively resolved as packets advance through the central modules and the implementation complexity is considerably reduced. In \cite{art15}, a HW implementation of a single-stage UDN packet switch is proposed where it is shown that a UDN module is perfectly feasible considering the current technology and that a cost/performance trade-off can be made by varying the switch parameters and/or the synthesis technology.

\subsection{Average blocking time in the IM queues}
We give a simplified estimation of the average waiting time that a packet spends in IM’s buffers under the following hypotheses. Input modules have Bernoulli uniform traffic arrivals occurring at a rate $\lambda$ and all queues operate independently one of the other with uniformly distributed service times. We denote $\frac{1}{\mu}$, the mean service time and $\rho=\frac{\lambda}{\mu}$, the traffic intensity. Given the above-mentioned approximations, we restrict the system analysis to a single FIFO with unlimited capacity behaving as an M/D/1 queue. Due to the feedback loop around the limited size of buffers of on-chip Mini-Routers (MRs), rise dependencies in the departure process from queues of the IMs. Packets are not dispatched to CMs unless the left-most buffers of the routers in the first column of the UDN have room. Next, we denote $BD$ the capacity of one buffer located at any of its ingresses. The probability that a packet gets forwarded to a CM can be expressed as:
\begin{equation}\label{eq:eq1}
	P_{fwd} = P_p~ (1-P_{ctr})
\end{equation}
where $P_p$ is the probability that a packet is present at the input FIFO and $ P_{ctr}$ is the flow-control probability issued by the left-most buffer of a MR to a requesting packet. 
Basically, a feedback-control signal is generated when a packet tries to access a saturated buffer which motivates the following expression for $P_{ctr}$:
\begin{equation}\label{eq:eq2}
	P_{ctr}= [P_p~ (1-P_{serv})]^{BD}
\end{equation}
$P_{serv}$ is the probability that a packet in one of the MR's input buffers is served and $n_{ingr}$ is the number of the MR's ingresses. Since a RR scheduling is adopted to fairly resolve contention among the MR's ingresses, $P_{serv}$ can be expressed as $P_{serv}=1/n_{ingr}$. Ultimately, we can approximate the average blocking delay a packet experiences in an input FIFO by:
\begin{equation}\label{eq:eq3}
	\omega_{block}= \frac{1}{2\mu^{\prime}}\left (\frac{\rho}{1-\rho}\right ) \
\end{equation}
where $\mu^{\prime}$ being the modified input queues service time given by:
\begin{equation}\label{eq:eq4}
	\mu^{\prime}= P_{fwd}~ \mu
\end{equation}


\section{Dealing with Out-Of-Sequence packets delivery}\label{Sec.out-of-seq}

\begin{figure*}
	\centering
	\includegraphics[width=4.3in]{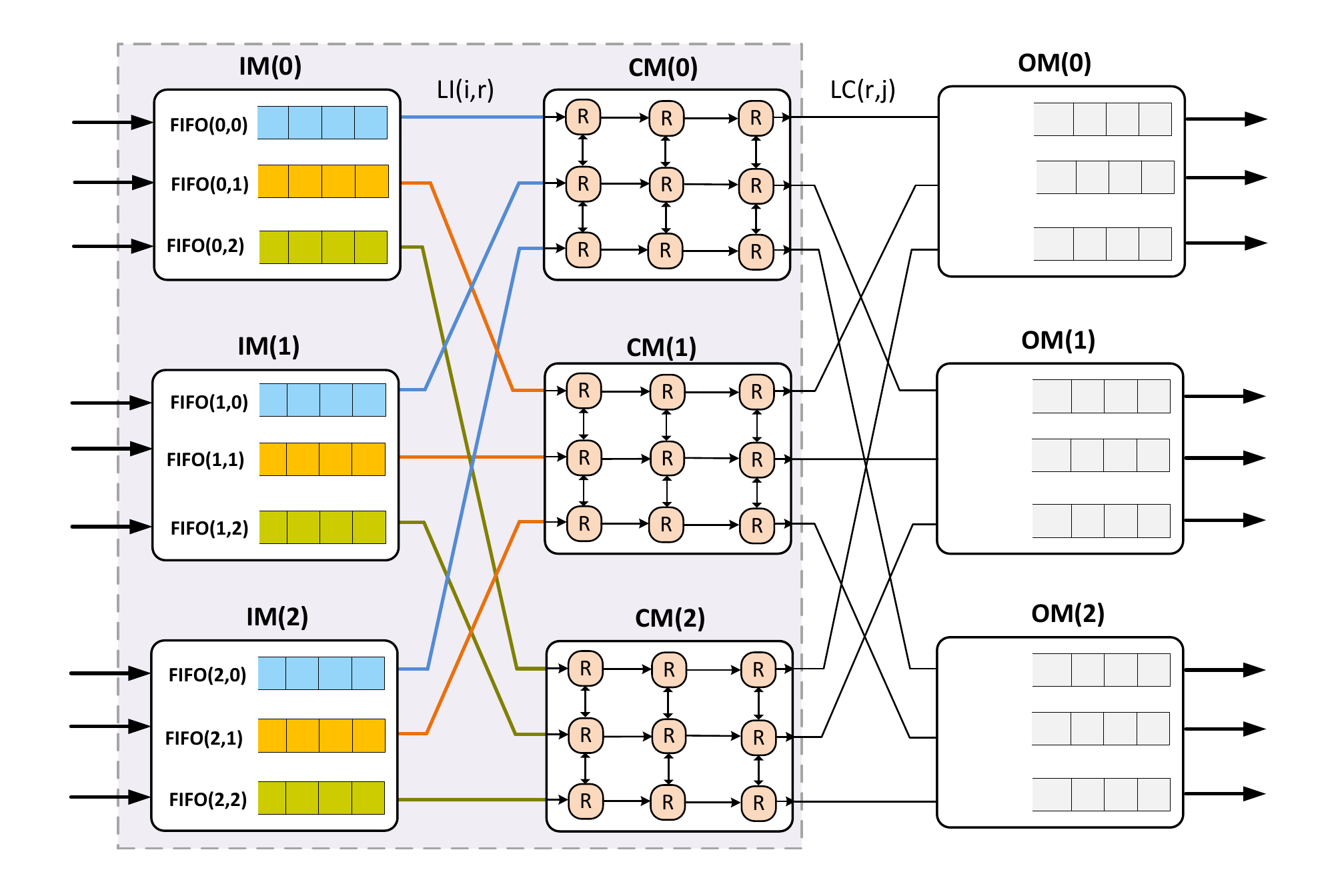}
	\caption{Example of a ($9\times 9$) Clos-UDN switch with static configuration of the IMs/CMs interconnections.
	} 
	\label{fig:modulo_shuffle} 
\end{figure*}

Pure space switching like S$^3$ introduces too much competition between cells at all stages. They suffer low throughput and have crippled performance. Introducing memories to the network switching fabric has always been the remedy to resolve contention. So far, buffreless architectures need complex and expensive matching mechanisms to make the path decision. However, they guarantee ordered packets delivery. In the opposite case, the fully buffered architecture MMM has good performance and simpler scheduling process. Yet, packets  experience variable queuing delays in the buffers leading in a mis-sequenced cells transmission \cite{art27}.

Ordering packets can be costly in terms of memory and processing time \cite{art7}\cite{art8}\cite{art17}. The proposed Clos-UDN with RR packet dispatching scheme suffers the same limitation. Packets are dynamically sent to the UDN central modules where they get serviced with variable delays. In the current work, we choose to alleviate packets from getting mis-sequenced in the first place. Based on the three-stage Clos-UDN switch architecture, we propose removing the $(n\times m)$ input schedulers and put in place a static configuration of the connections between the input FIFOs of IMs and CMs' ingresses (TABLE \ref{tab:tab_3}).
\begin{table}[!htbp]
	\centering
	\setlength{\tabcolsep}{.3em} 
	{\renewcommand{\arraystretch}{1.2}
		\caption{HW requirements: Dynamic vs static dispatching in Clos-UDN switch} \label{tab:tab_3} 
		\begin{tabular*}{\columnwidth}{c| c| c| c} \cline{2-3}
			& Three-stage Clos-UDN & Two-stage Clos-UDN & \\ \cline{1-3}
			\multicolumn{1}{|c|}{Dispatching scheme} &   Dynamic RR dispatching  & Static dispatching   & \\ \cline{1-3}
			\multicolumn{1}{|c|}{IM arbiters} & $m$ & NA & \\ \cline{1-3}
			\multicolumn{1}{|c|}{In-order packets delivery} & No & Yes & \\ \cline{1-3}
			\multicolumn{1}{|c|}{Performance / scalability  } &    High  & Good  & \\ \cline{1-3} 
		\end{tabular*}	
	}
\end{table}

Consequently, the new switching architecture can be viewed as a two-stage network. The intuition behind the two-stage approach is as follows. It is known that the $"Modulo~XY"$ algorithm used to route packets inside the UDN modules is a deterministic minimal paths algorithm that ensures that packets of the same flow do follow the same path throughout the NoC fabric. Imposing a static dispatching scheme effectively makes traffic flows routed in order through the whole Clos-UDN network. Unlike previous proposals, no re-sequencing buffers, synchronization signals and complex algorithms are required. Theoretically, a two-stage Clos-network is only rearrangeably non-blocking. As path relocation is prohibited in practical Clos switches, two-stage architecture becomes blocking and compromises throughput \cite{art29}. All the same, two-stage interconnects prove to be interesting mainly for optical switching architectures \cite{art7}\cite{art32}\cite{art35}. The architecture improvements from the single-stage baseline are still significant. Although, the two-stage Clos-UDN diminishes the extra switching area per IM/OM pair, it contributes toward reducing the cost/complexity and notably ensuring an ordered packets delivery. The switch is still scalable thanks to the scalable central modules and presents good complexity and performance trade-off for high-radix switches.

\section{Performance evaluation}\label{Sec.Sim}
We use an event-driven simulator to evaluate the performance of the three-stage Clos-UDN with a dynamic and a static dispatching schemes. Unless it is mentioned, the switch size is set to ($64\times64$), the on-chip buffers ($BD$) are worth of four packets each and square UDN meshes are considered for the central stage of the Clos-network $(M=k)$. The simulation time is $10^6$ time slots. 
In the simulation figures, $iterx$ stands for $x$ number of iterations for the CRRD dispatching algorithm and $SPx$ means that the on-chip links of the UDN modules in the Clos-UDN switch run $x$ times faster than the LI/LC links. As for MMM, $xbuff$ is the size of the middle stage's crosspoint buffers.

\begin{figure*}[htbp]
	\minipage{0.46\textwidth}
	\includegraphics[width=\linewidth]{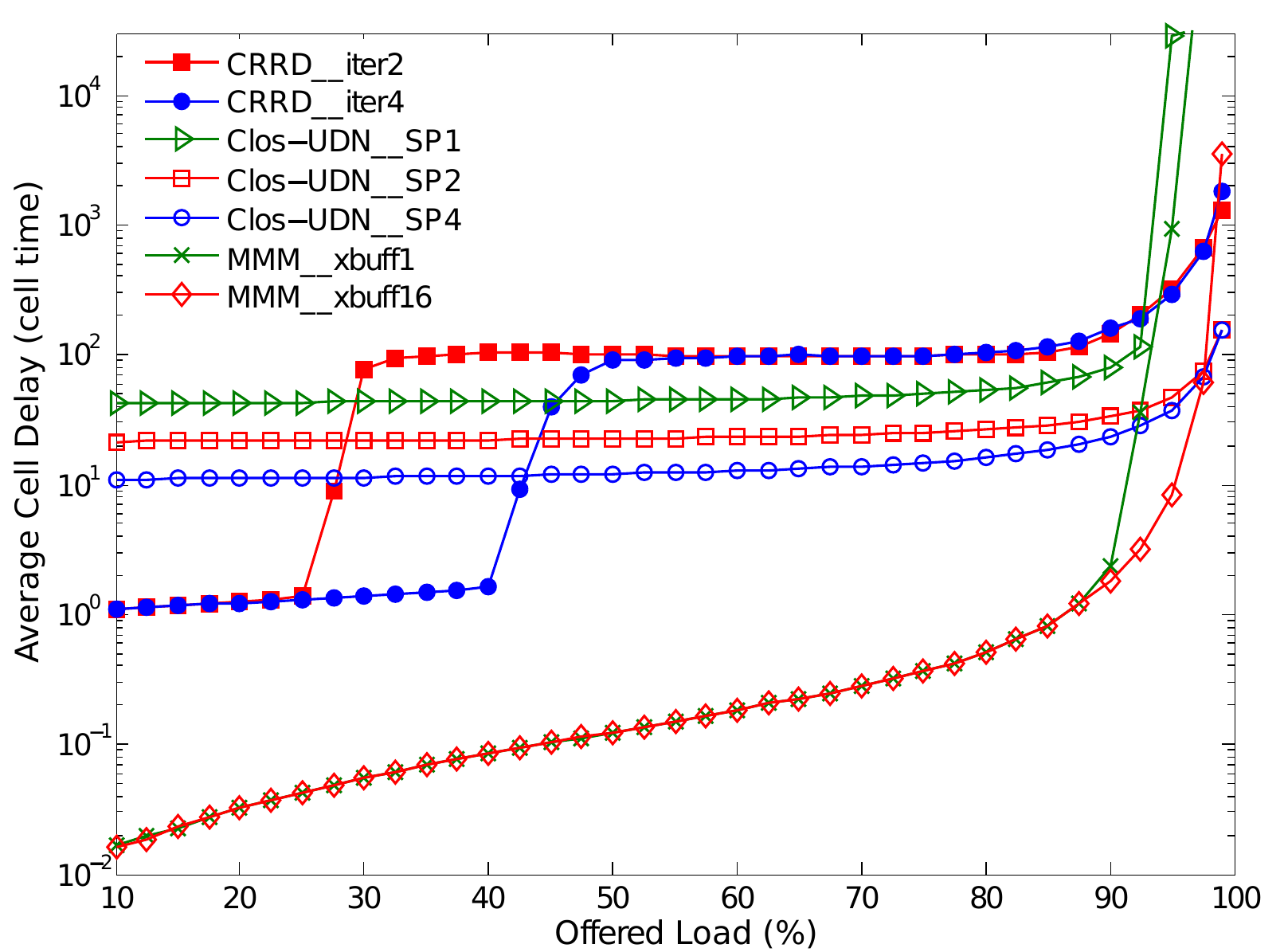}\vspace{.3cm}
	\caption{Delay performance of Clos-UDN, MMM and MSM \\ using CRRD algorithm, Switch size=256, Bernoulli uniform \\ traffic}\label{fig:compare}
	\endminipage\hfill
	\minipage{0.48\textwidth}%
	\includegraphics[width=\linewidth]{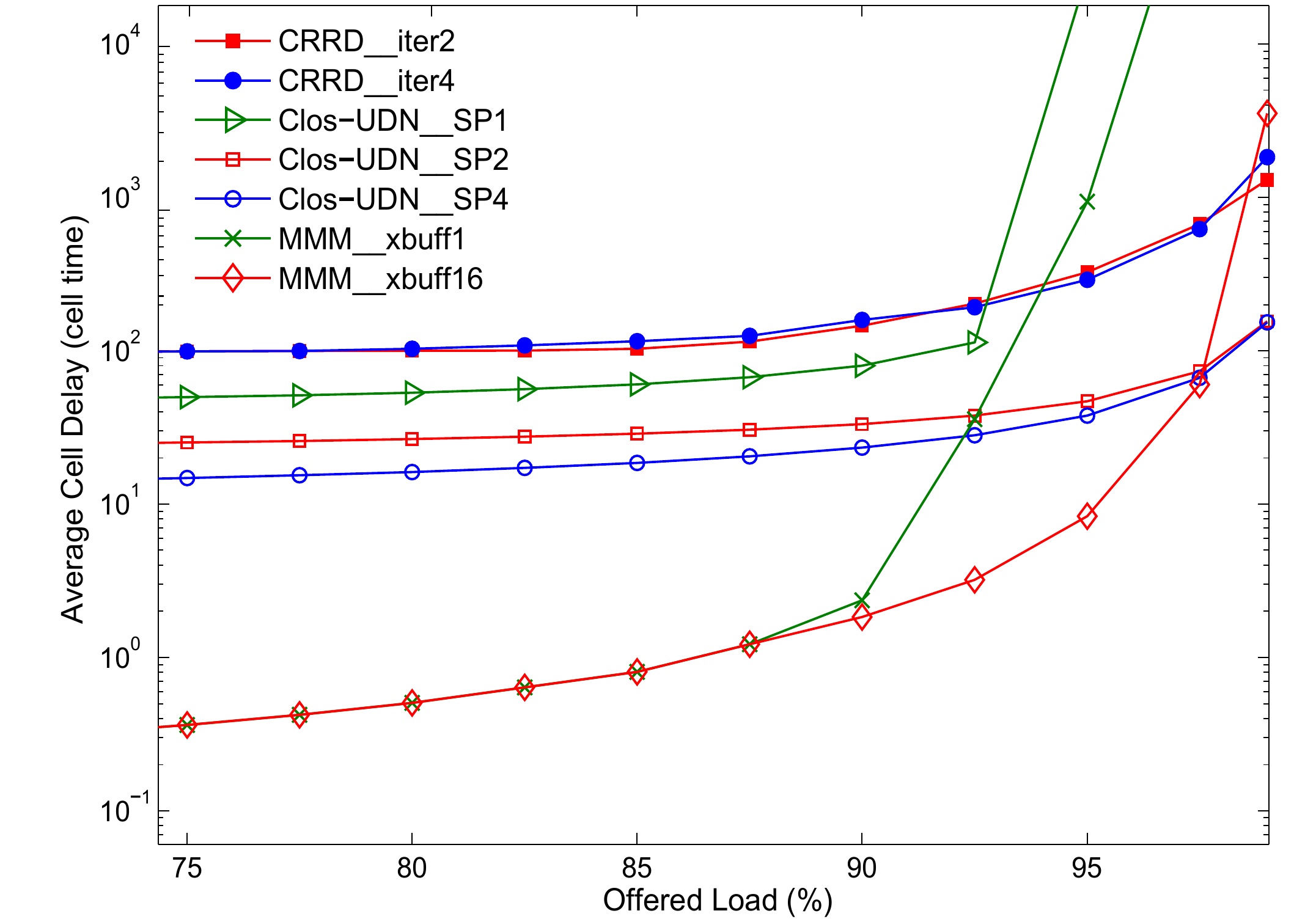}\vspace{.3cm}
	\caption{Zoomed view of Fig.~\ref{fig:compare} - Delay performance under high Bernoulli uniform traffic loads}\label{fig:compare_zoom} 
	\endminipage
\end{figure*}

\begin{figure*}[htbp]
	\minipage{0.46\textwidth}
	\includegraphics[width=\linewidth]{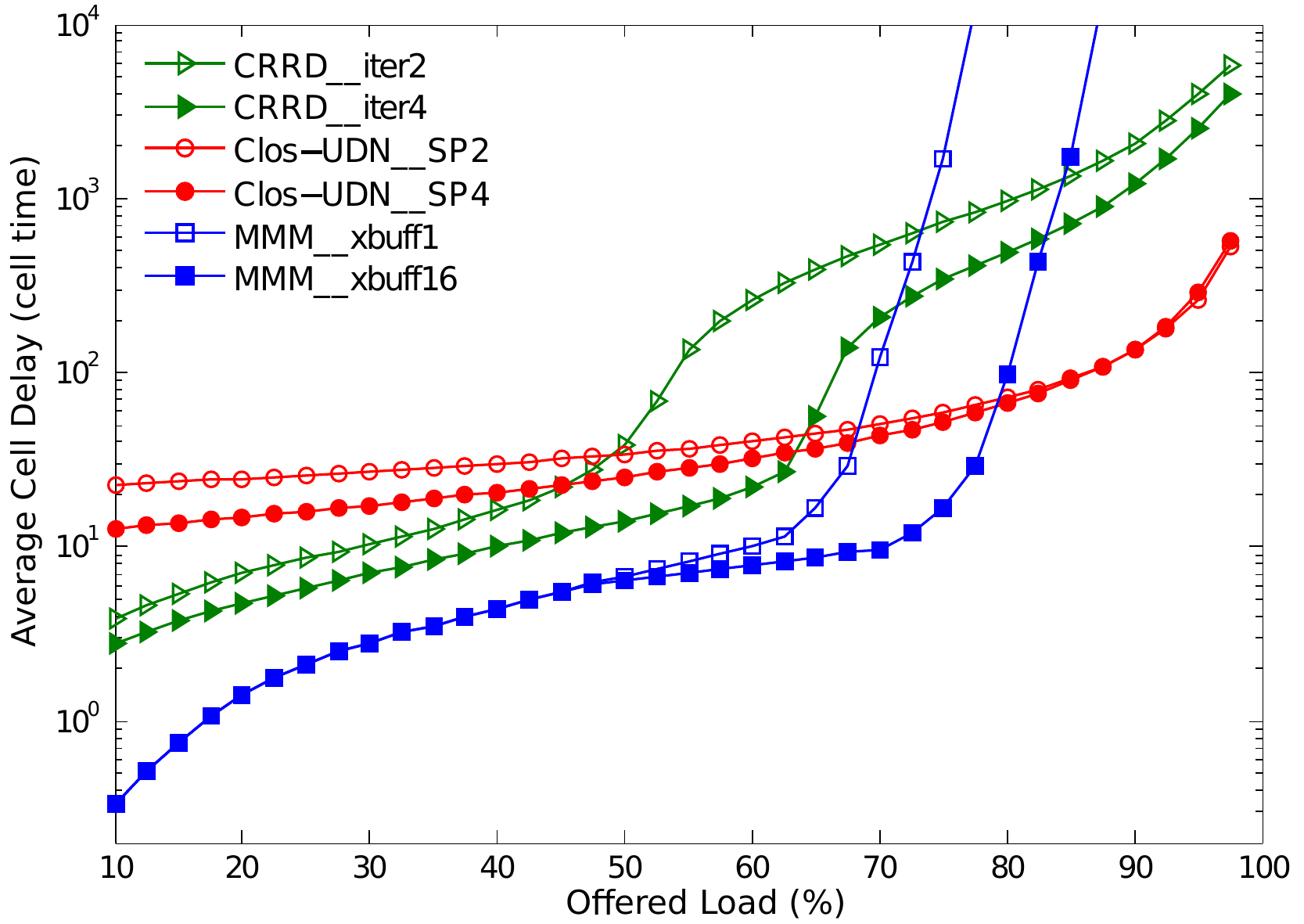}\vspace{.3cm}
	\caption{Delay performance of Clos-UDN, MMM and MSM using CRRD, Switch size=256, Bursty uniform traffic}
	\label{fig:burst}
	\endminipage\hfill
	\minipage{0.46\textwidth}%
	\includegraphics[width=\linewidth]{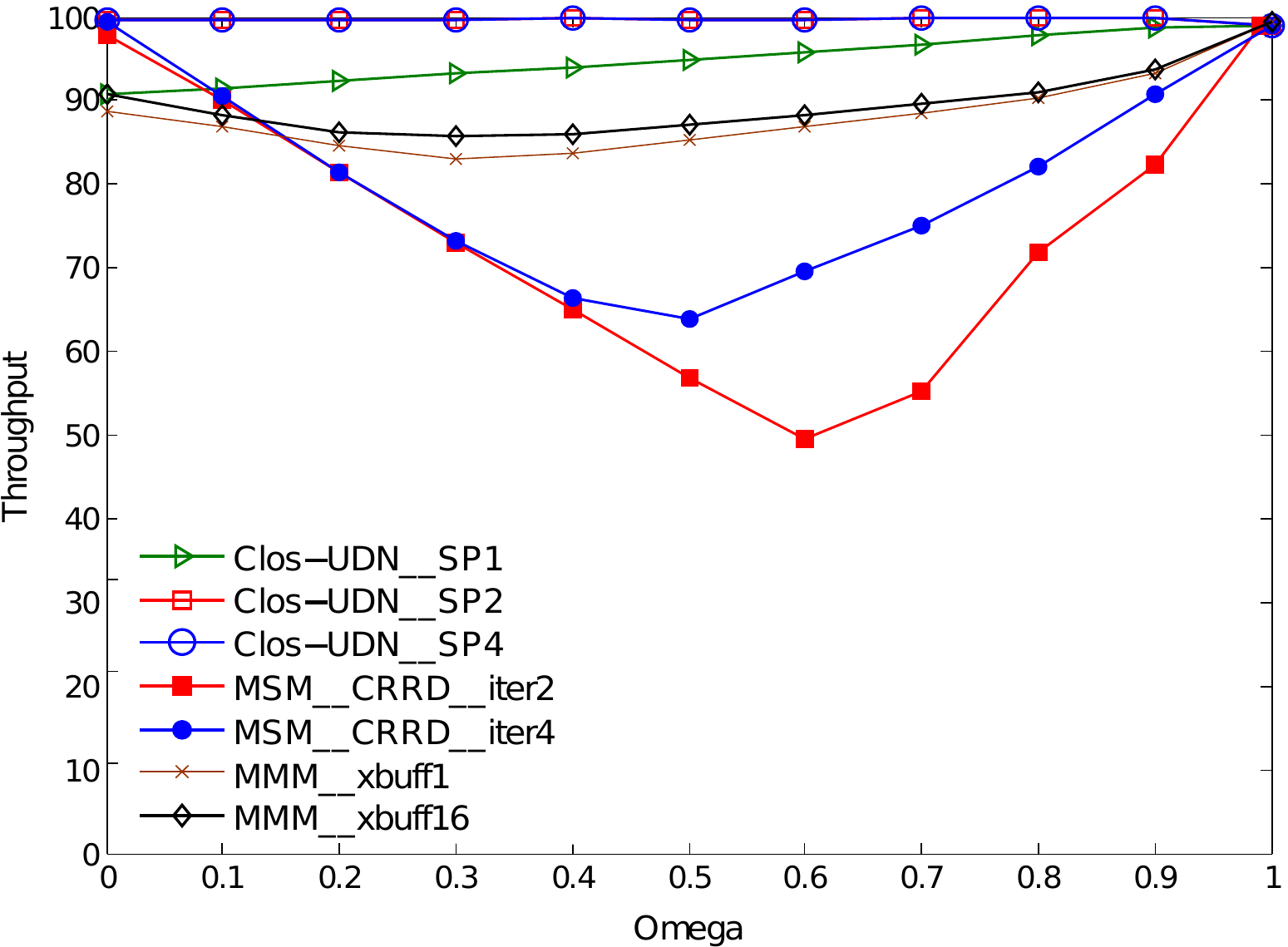}\vspace{.3cm}
	\caption{Throughput stability of Clos-UDN, MSM and MMM, Unbalanced traffic}
	\label{fig:throughput}
	\endminipage
\end{figure*}

We compare the performance of the Clos-UDN switch to the MSM switch architecture with CRRD dispatching as described in \cite{art4} (baseline) and MMM as discussed in \cite{art7}. Both architectures are compared under various switch sizes and traffic scenarios settings including uniform Bernoulli and bursty as well as non-uniform traffic arrivals. 
It is important to note that the speedup used in the Clos-UDN is different than the conventional speedup~\cite{speedup}, where $SP$ refers to the internal switch over-speed factor with respect to the external line rate. Here $SP$ refers to the over-speed factor of \textit{only} the NoC routers inside each UDN central module, excluding the LIs and LCs. Meaning, just like the MSM, the Clos-UDN always sends at maximum one packet per LI/LC link per time slot. Since the Clos-UDN does not use iterations (i.e. time) in its matching, this could be compensated by internally running the UDN CMs with small $SP$ values.


\subsection{Clos-UDN vs. MSM and MMM}
\subsubsection{Bernoulli uniform traffic}

Fig.~\ref{fig:compare} compares the packets delay performance of the Clos-UDN  using different speedup values to that of MMM switch and MSM switch employing CRRD with different iterations. We simulate the Clos-UDN switch with 4-packets large buffers at the on-chip routers' inputs making a total of 12 packets buffering per crosspoint. 


\begin{figure*}[htbp]
	\minipage{0.45\textwidth}
	\includegraphics[width=\linewidth]{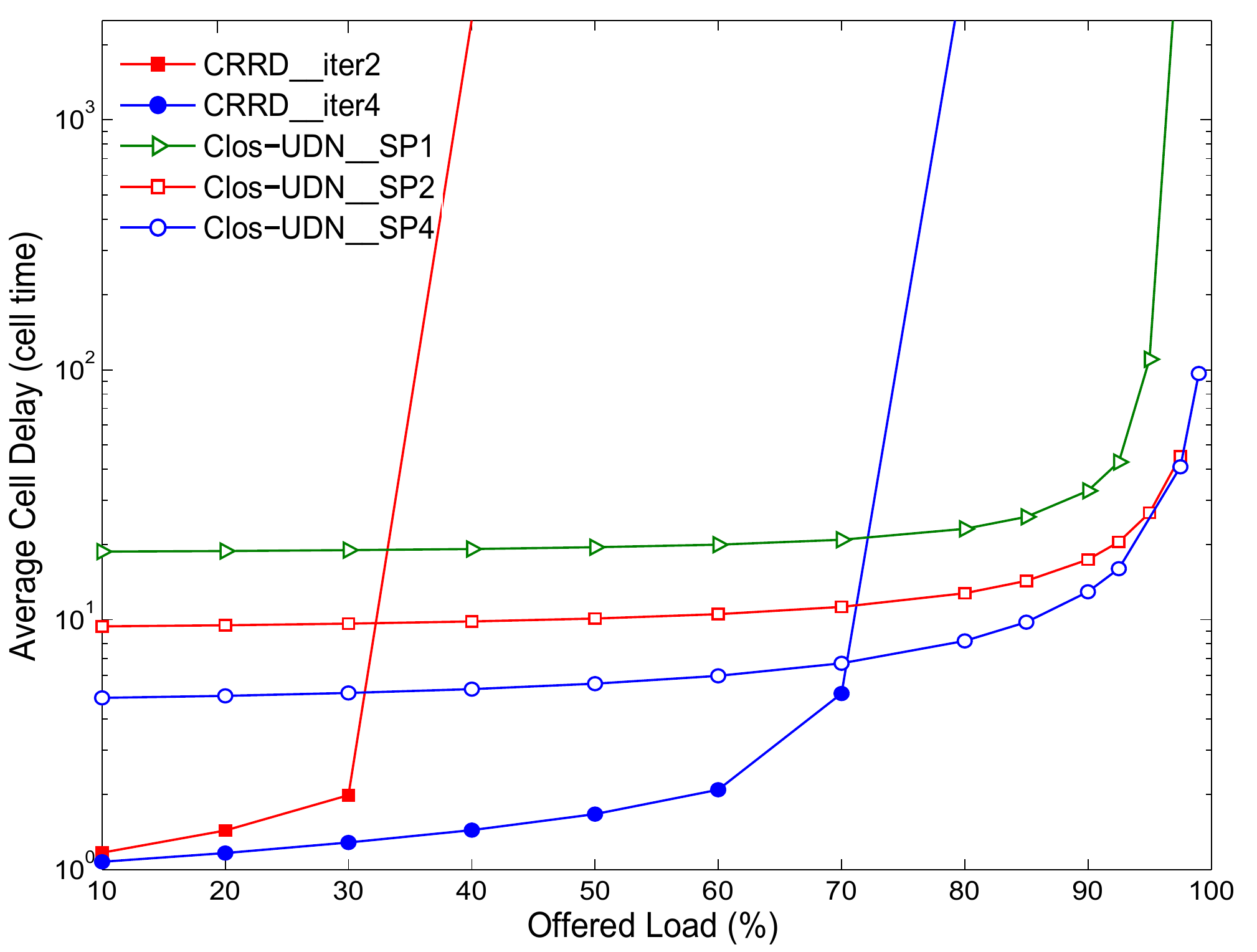}\vspace{.3cm}
	\caption{Delay Performance of Clos-UDN and MSM, Hot-Spot traffic}
	\label{fig:unbalanced}
	\endminipage\hfill
	\minipage{0.46\textwidth}%
	\includegraphics[width=\linewidth]{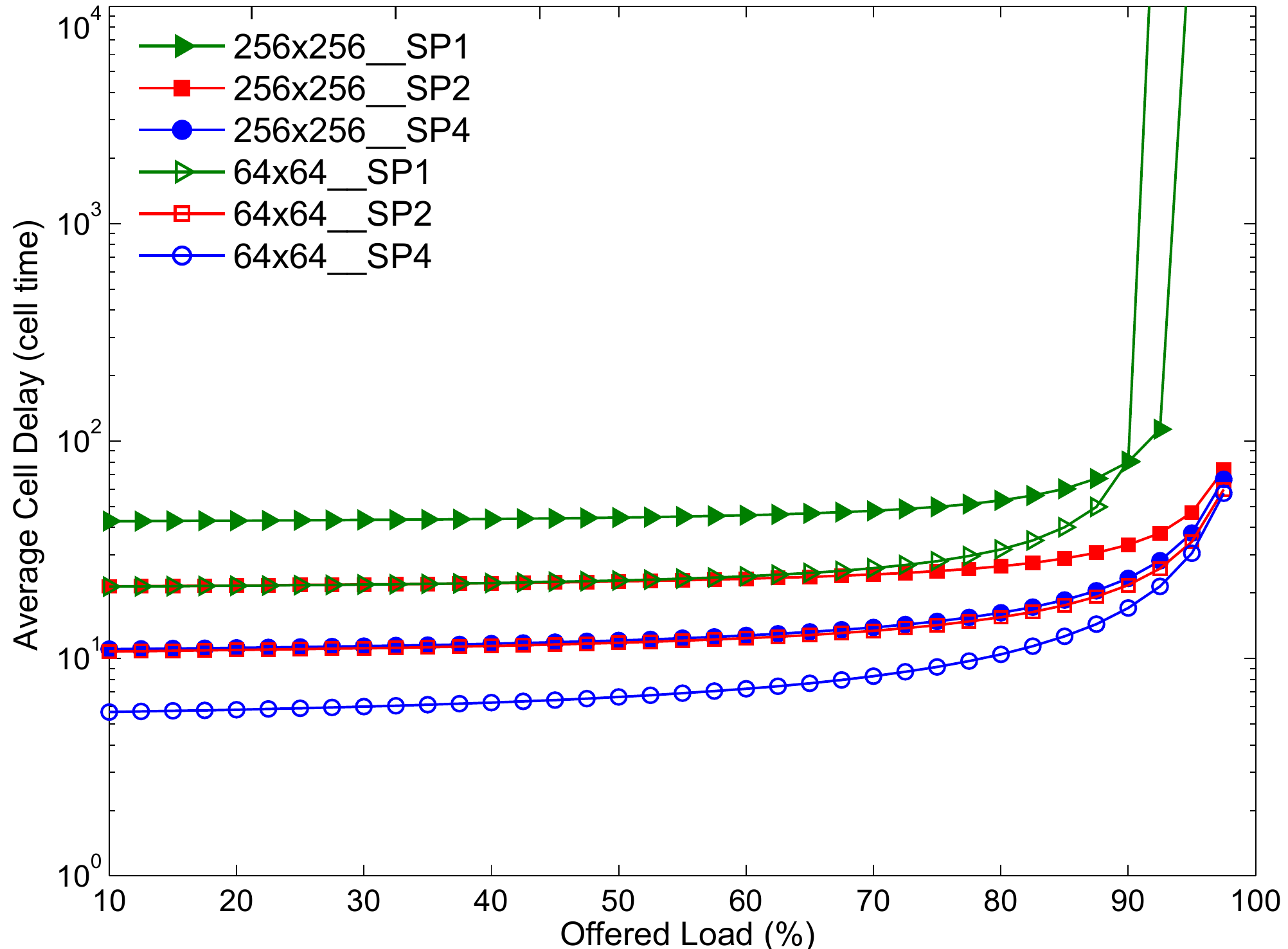}\vspace{.3cm}
	\caption{Impact of the switch size and speedup on the delay performance of the Clos-UDN, Bernoulli Uniform traffic}
	\label{fig:switch_size}
	\endminipage
\end{figure*}

Running the CM units at a speedup factor of one makes the switch achieve 90${\%}$ throughput. It is the packets progressing in the central switching units by one at each cycle ($SP=1$) that prevents the switch from achieving full throughput.  Increasing the speedup factor to two suffices for the switch to achieve full-throughput. The proposed switch architecture outperforms the MSM under medium-to-high uniform traffic arrivals (which are more relevant in the context of DCNs). The slightly higher delay experienced by the Clos-UDN under light loads is due to the time required to fill-in the pipeline of the multi-hop NoC based CMs as shown in Fig.~\ref{fig:pipelining}. However, Clos-UDN maintains low and almost constant delay irrespective of the traffic load. When the load is larger than 0.9, the delay performance of Clos-UDN with $SP=2$ becomes better than MSM using CRRD with 4 iterations and MMM as Fig.~\ref{fig:compare_zoom} shows. We note that increasing the speedup of UDN switches pulls down the cell delay. MMM behavior approaches an OQ switch. It outperforms MSM and Clos-UDN switches if light to medium loads strike the switch inputs. Setting the crosspoint buffers' size to $b=1$, makes the MMM throughput almost equal to $94\%$ and full throughput is achieved if the buffers are as large as $b=16$. Clearly, Clos-UDN deals in a better way with high loads for which it keep the lowest system delay (compared to MSM and MMM, for high loads).

\subsubsection{Bursty uniform traffic}
High-bandwidth demanding applications make the bursty traffic pattern prevalent in a data center network with high-levels of peak utilization. We examine the effect of burstiness on the Clos-UDN switch by considering a bursty traffic with a default burst length equal to 10. Fig.~\ref{fig:burst} reveals that the delay's growth of the Clos-UDN under bursty arrivals is smoother than that of the MSM with CRRD even if the matching procedure runs 4 iterations. Increasing the number of iterations for the CRRD provides better matching between IMs and CMs and resolves faster the contention which lead to improved switch performance when the load is below 0.7. Increasing the $\it{SP}$ reduces the initial delays for the Clos-UDN. Simulations show that MMM is less efficient when evaluated for bursty traffic. Although the switch has lower average packet delay, Clos-UDN with a minimum $SP=2$ proves to outperform both MSM and MMM under heavy bursty traffic arrivals. Visibly, MMM has degraded throughput and increasing the crosspoint  buffers to $b=16$ is of little effect as it only shifts the switch throughput from $77\%$ to $86\%$.



\subsubsection{Unbalanced traffic} 
Next, we evaluate the Clos-UDN switch under non-uniform  unbalanced traffic, as specified in \cite{art4}. This traffic pattern has one fraction of the total load generated uniformly and the other fraction destined to the output with the same index as the issuing input. If $\omega=0$, then the traffic is perfectly uniform. If $\omega = 1$, the switch deals with a totally unbalanced traffic. We evaluate the throughput performance of the proposed switch. We reproduce the results for the MMM as described in \cite{art7} where the Longest Queue First (LQF) selection at the input ports and RR arbiters in the different modules are used. Fig.~\ref{fig:throughput} depicts the switch throughput when we vary the unbalancing coefficient $\omega$.  The Clos-UDN with $SP=1$ achieves 90${\%}$ throughput for $\omega = 0$ (uniform traffic), as has been already shown in Fig.~\ref{fig:compare}. MMM achieves better throughput than MSM switch performing CRRD scheduling (60${\%}$ throughput if 4 iterations are used and $\omega=0.5$). However Clos-UDN has higher and more stable throughput variation than both semi-buffered and fully buffered architectures under the whole range of $\omega$. 

Setting $\omega = 0.5$ corresponds to a non-uniform hot-spot traffic, where 50\% of the input load goes to one output while the rest is equally distributed over the remaining outputs. A further step in analyzing the Clos-UDN switch performance consists on inspecting the average delay under non-uniform traffic pattern in comparison to MSM. Fig.~\ref{fig:unbalanced} presents the  results, under these settings,  for a ($64\times 64$) switch operating both the Clos-UDN and MSM architectures. Curves in Fig.~\ref{fig:unbalanced} point out that the Clos-UDN switch architecture has much better average delay than the MSM, irrespective of the Clos-UDN speedup and the CRRD number of iterations.

\subsection{Further analysis of the Clos-UDN switch}
In this subsection we vary a set of parameters of the Clos-UDN and study the effect of each one on the overall switch'performance.
\vspace{.3cm}
\subsubsection{Varying the switch size}
Performance curves depicted in Fig.~\ref{fig:switch_size} show that increasing the Clos-UDN valency has a minor effect of the overall delay, making it truly a scalable solution and a good alternative for DCN Top-of-Rack switches. Large switches can achieve good performance if the $\it{SP}$ is increased to just 2. A ($256\times 256$) Clos-UDN switch running at a speed  $SP=4$ has an average cell latency that is approximately the same as ($64\times 64$) switch with  $SP=2$. 


\begin{figure*}[htbp]
	\minipage{0.46\textwidth}
	\includegraphics[width=\linewidth]{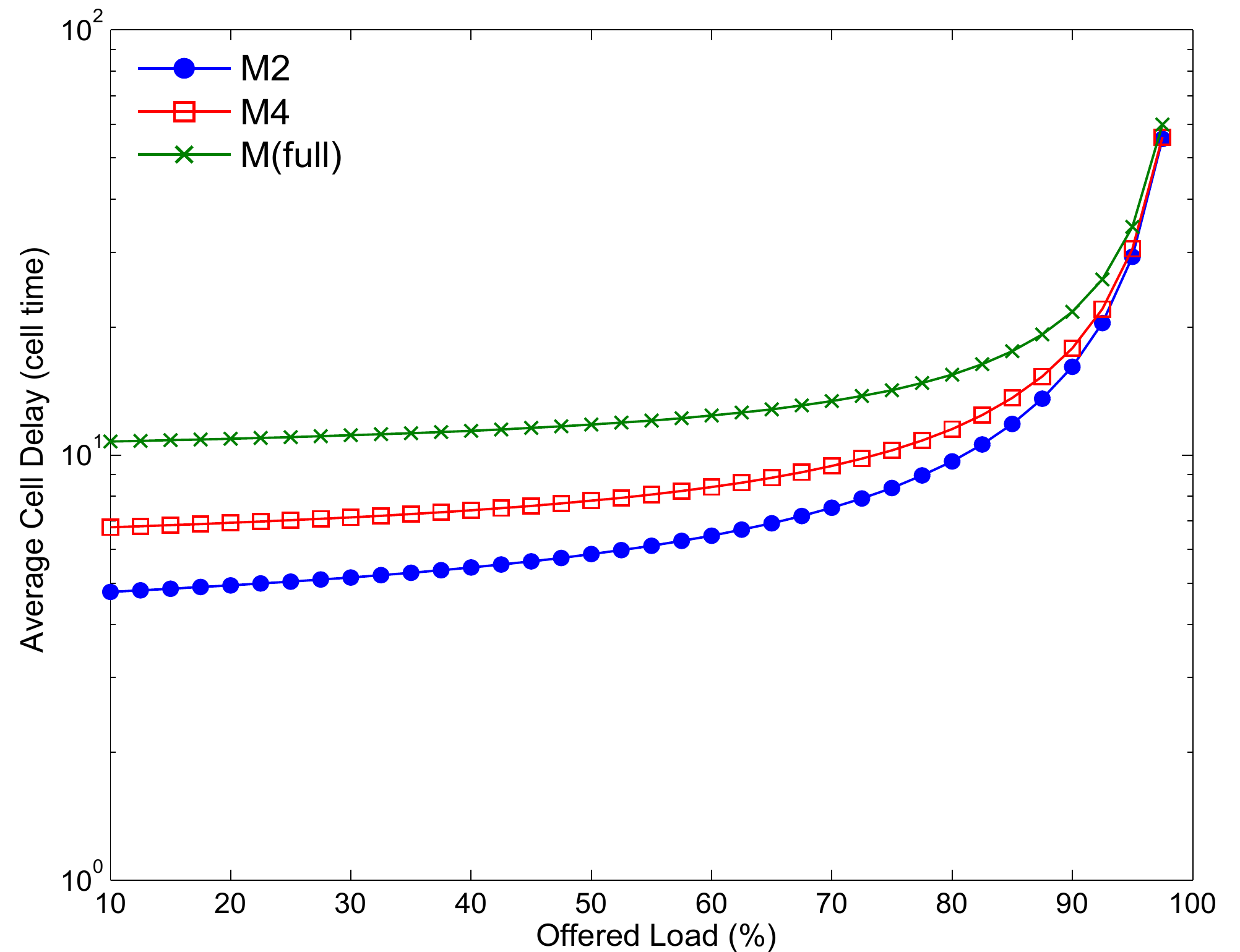}\vspace{.3cm}
	\caption{Impact of the UDNs' mesh depth on the delay \\performance of the Clos-UDN switch, Bernoulli Uniform traffic, $SP=2$}
	\label{fig:depth}
	\endminipage\hfill
	\minipage{0.46\textwidth}%
	\includegraphics[width=\linewidth]{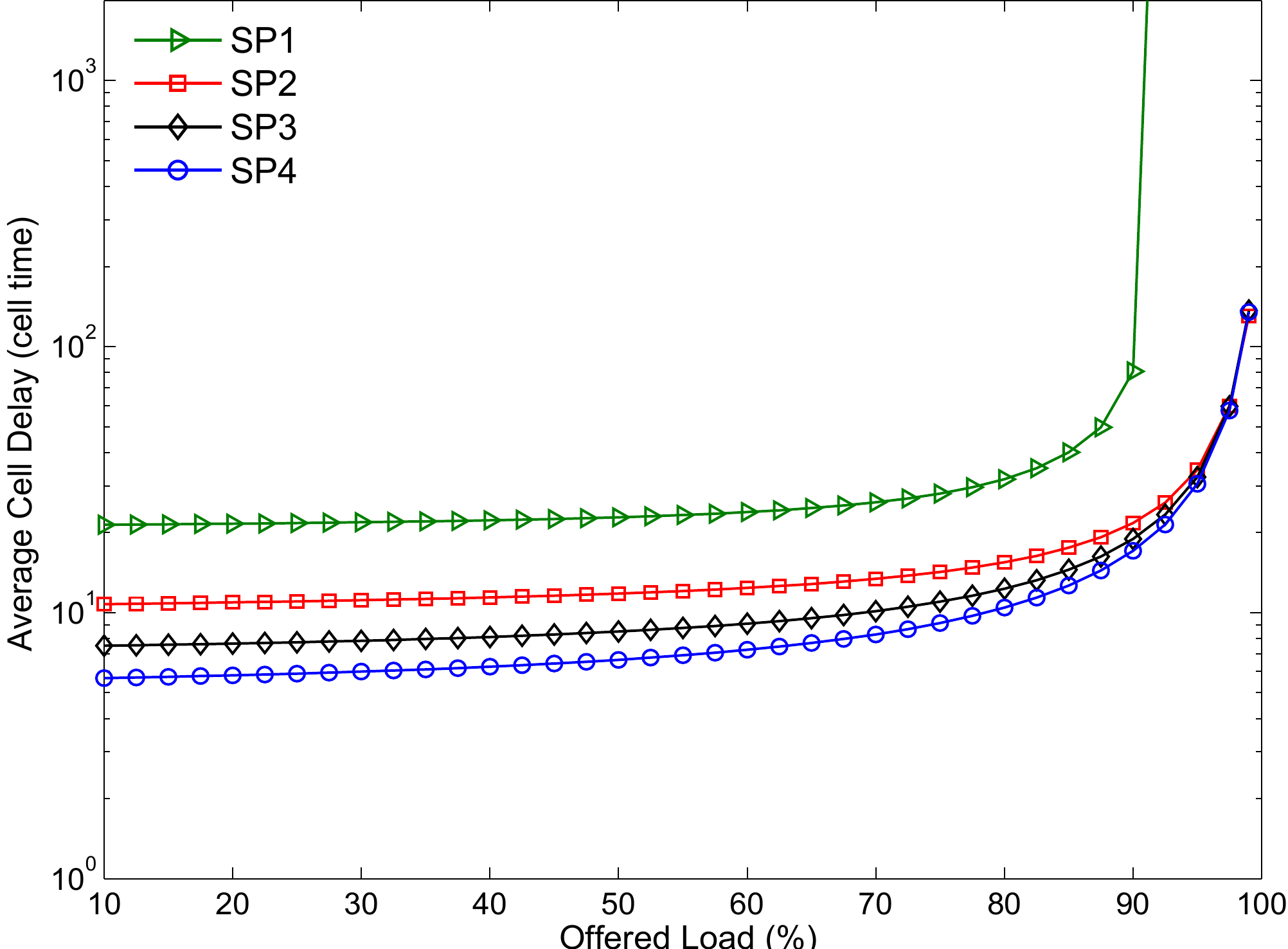}\vspace{.3cm}	
	\caption{Impact of the speedup on the delay performance of the Clos-UDN switch, Bernoulli traffic}
	\label{fig:speed}\label{fig:clos}
	\endminipage
\end{figure*}

\vspace{.2cm}
\subsubsection{Varying the depth of the UDN units}~\label{udn.depth}
The Clos-UDN is configurable. Changing the number of the UDN's intermediate stages ($\it{M}$) can be done to trade-off cost/performance\cite{art13}\cite{art15}. However, this cannot be done without limits as it may cause the structure of the NoC to be congested and the performance to collapse. The Clos-UDN's initial latency is acquired from the multi-hop nature of the NoC-based CMs. In conventional crossbars, packets cross the fabric in one shot. However, in UDN, they have to cross at least  $\it{M}$ on-chip routers to reach their destination which results in a cumulative delay. Reducing $\it{M}$ causes the packets to travel through less intermediate stages before arriving to LC links. Hence, the average packets latency gets low when the switch is non-congested for fabrics running at a minimum  $SP=2$ as Fig.~\ref{fig:depth} shows.


\begin{figure*}[htbp]
	\minipage{0.47\textwidth}
	\includegraphics[width=\linewidth]{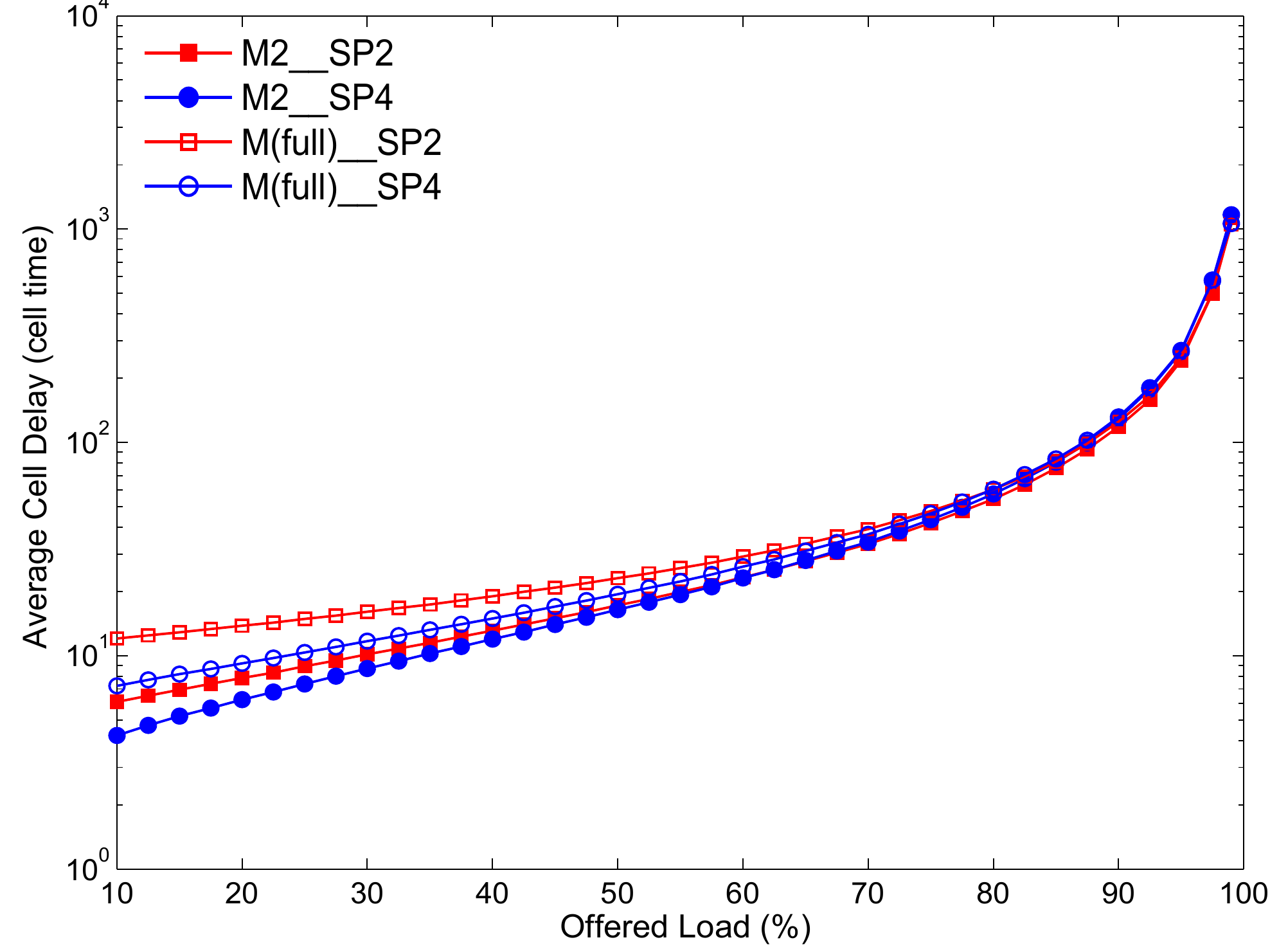}\vspace{.3cm}	
	\caption{Impact of the speedup $SP$ and mesh depth $M$ on the Clos-UDN switch latency, Bursty traffic }
	\label{fig:speed_depth}
	\endminipage\hfill
	\minipage{0.46\textwidth}%
	\includegraphics[width=\linewidth]{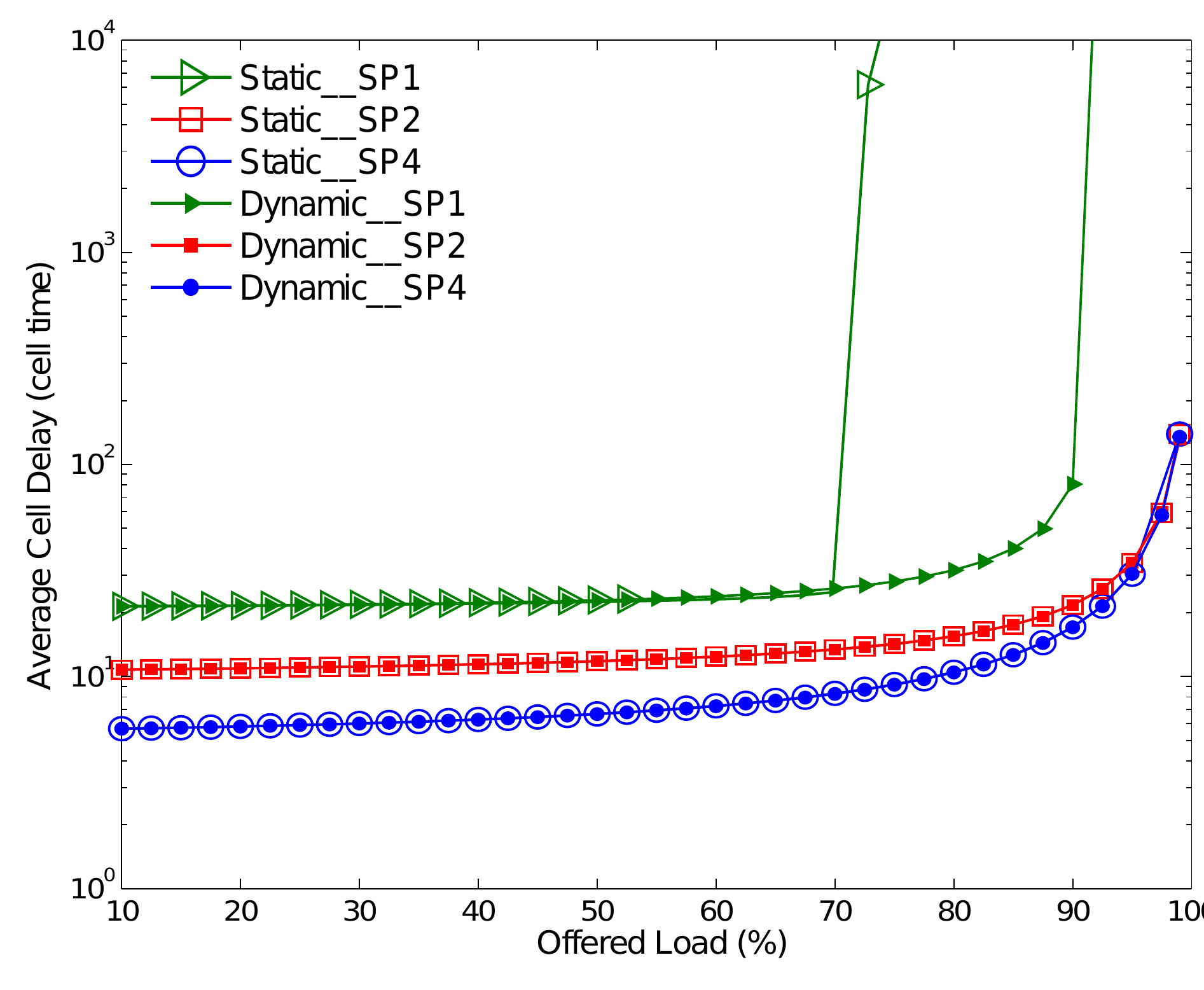}\vspace{.3cm}
	\caption{Delay performance of the two-stage Clos-UDN switch, Bernoulli uniform traffic}
	\label{fig:shuffle_uniform_traffic}
	\endminipage
\end{figure*}

\vspace{.2cm}
\subsubsection{Running the central modules faster}
As Part of the Clos-UDN architecture, all embedded routers of the CM blocks are input-buffered routers that require speedup to achieve full-throughput. Fig.~\ref{fig:speed} shows that increasing $\it{SP}$ contributes toward diminishing the overall latency when the switch is less congested. Reducing the number of columns $\it{M}$ to only two improves the system\textquotesingle s latency under light traffic loads. Fig.~\ref{fig:speed_depth} shows that decreasing $\it{M}$ is more effective than running the CMs faster as (increasing $\it{SP}$ for a given depth $\it{M}$). We conclude that running UDN central modules faster for a given mesh depth has less impact than dropping ($\it{k}$-$\it{M}$) columns and that one can choose the best settings for the Clos-UDN architecture to achieve pre-estimated performance levels. 

\vspace{.2cm}
\subsubsection{Changing the Buffer Depth }
A single router in a UDN module is a complete switching element with small input memories and a processing unit. Input buffers account for the major part of a router\textquotesingle s area which needs to be as small as possible for cost saving reasons. Increasing the $BD$ improves the system latency, but reducing the buffering amount can produce problems related to insufficient buffering \cite{art5}.

\subsection{Performance of the two-stage Clos-UDN switch}



\begin{figure*}[htbp]
	\minipage{0.45\textwidth}
	\includegraphics[width=\linewidth]{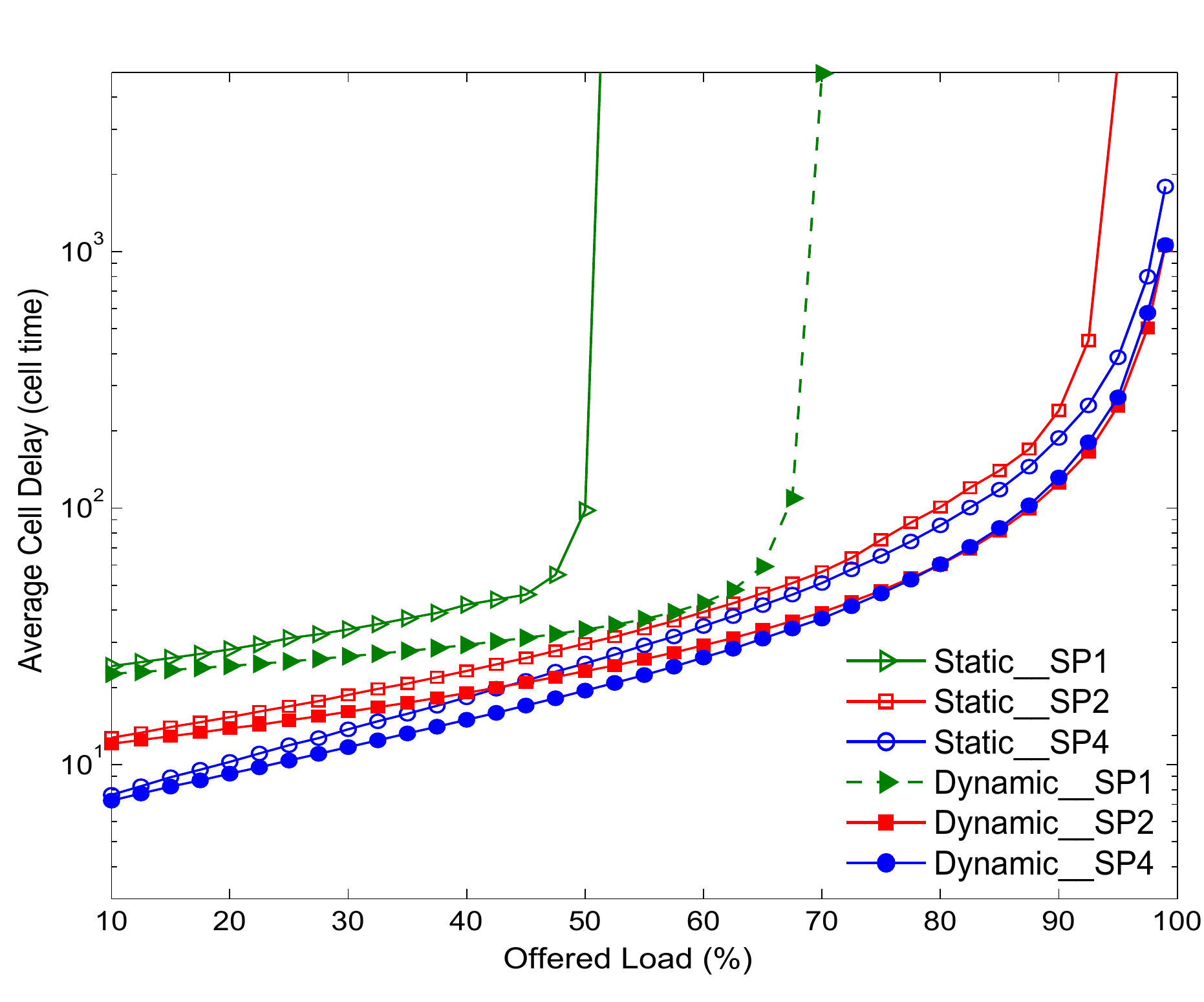}\vspace{.3cm}	
	\caption{Delay performance of the two-stage Clos-UDN switch, Bursty uniform traffic}
	\label{fig:shuffle_bursty_traffic}
	\endminipage\hfill
	\minipage{0.468\textwidth}%
	\includegraphics[width=\linewidth]{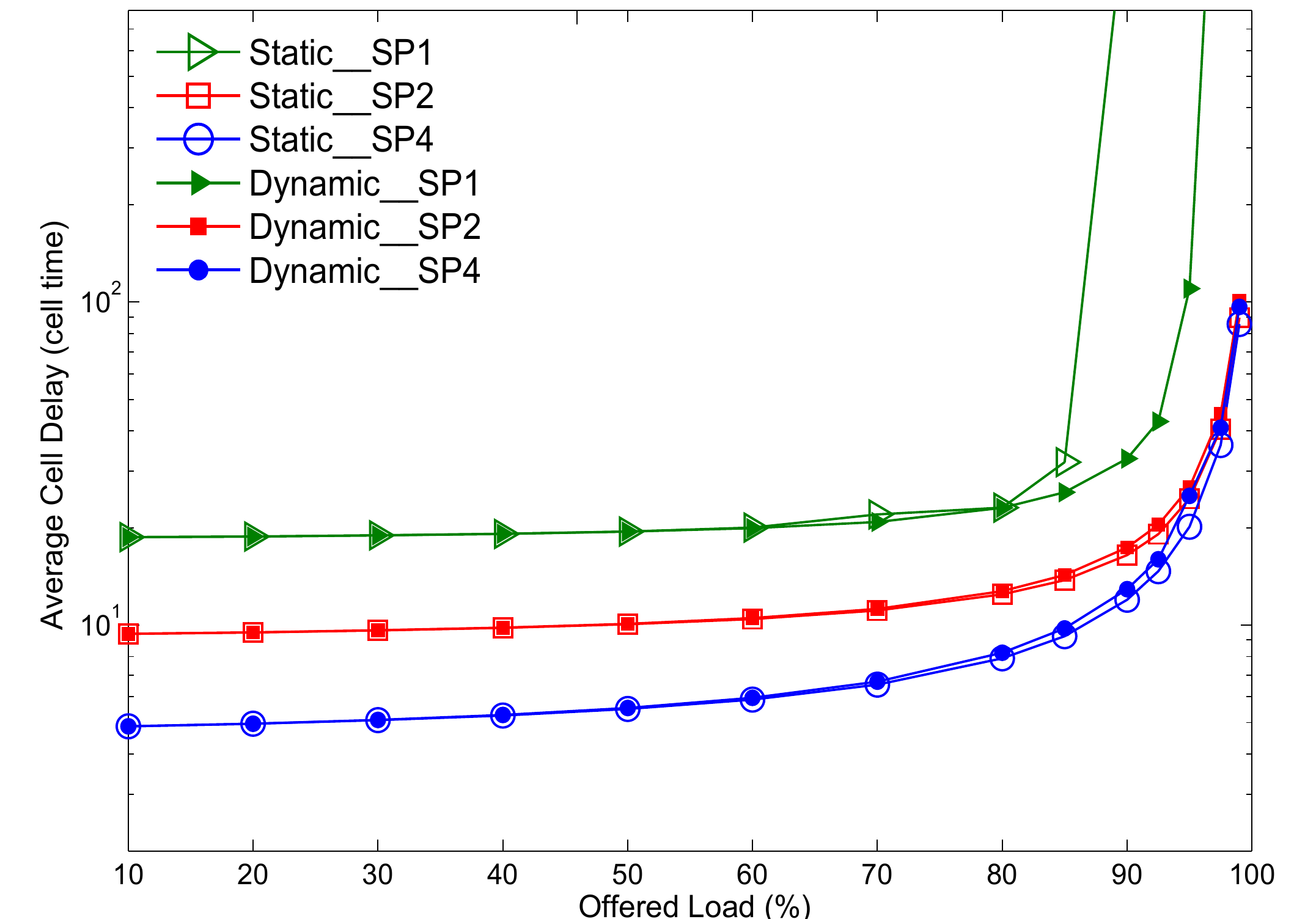}\vspace{.3cm}
		\caption{Delay performance of the two-stage Clos-UDN switch, Hot-Spot traffic}
		\label{fig:shuffle_unbalanced_traffic}
	\endminipage
\end{figure*}

\begin{figure*}[htbp]
	\minipage{0.46\textwidth}
	\includegraphics[width=\linewidth]{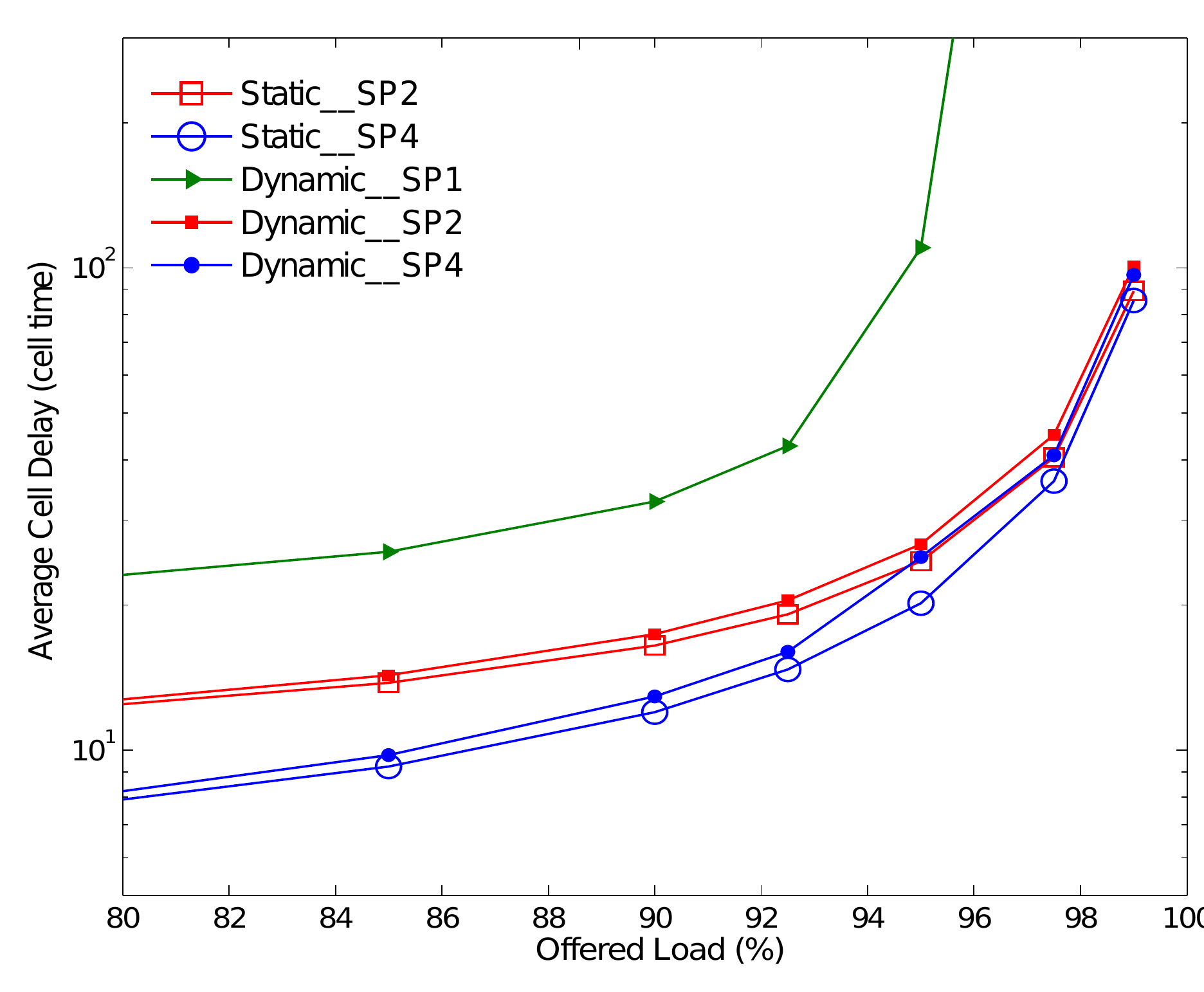}	\vspace{.3cm}	
	\caption{Zoomed view of Fig.~\ref{fig:shuffle_unbalanced_traffic} - Delay performance under high Hot-Spot traffic loads }
	\label{fig:zoom_unbal}
	\endminipage\hfill
	\minipage{0.46\textwidth}%
	\includegraphics[width=\linewidth]{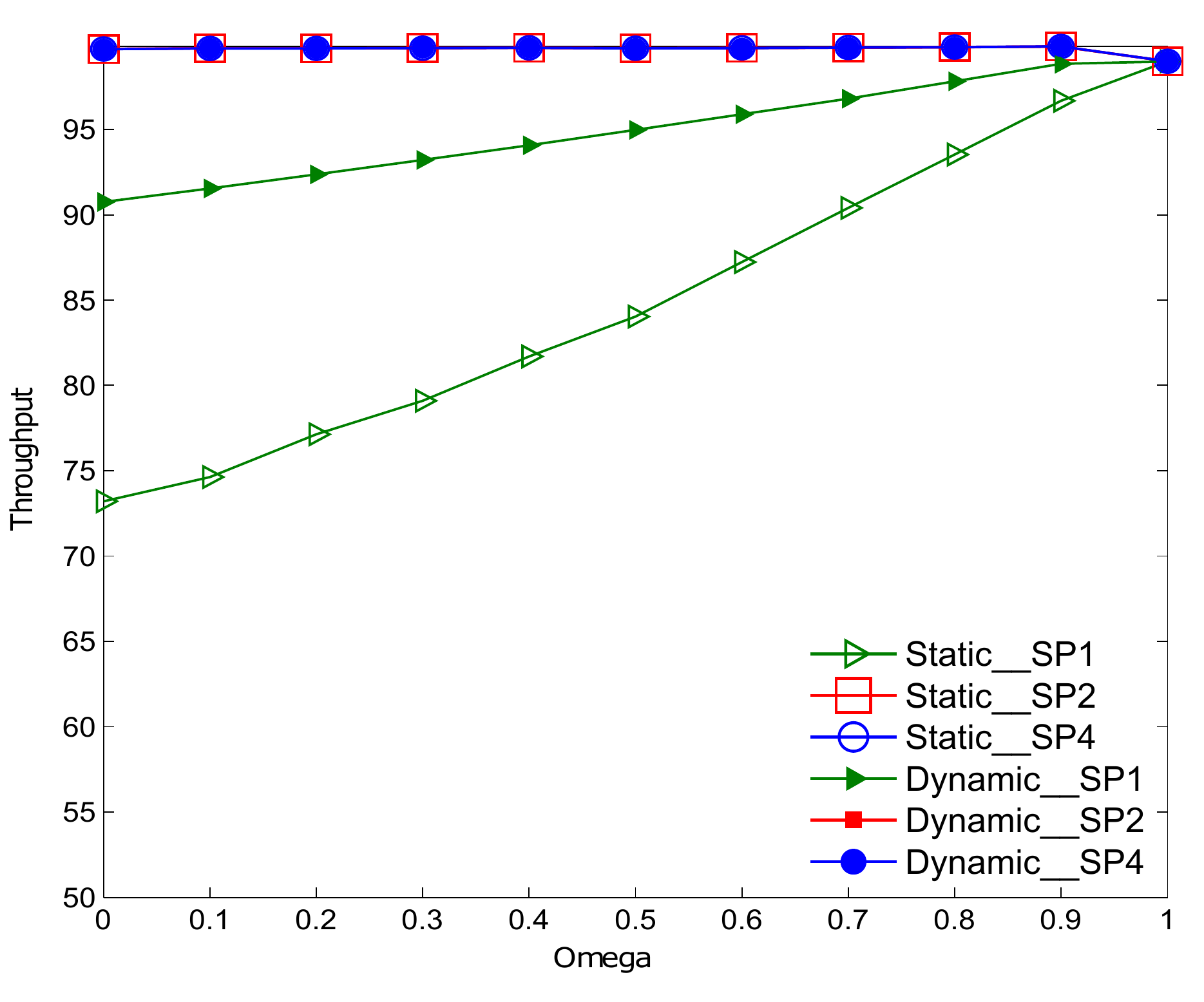}	\vspace{.3cm}	
	\caption{Throughput stability of the two-stage and three-stages Clos-UDN under Unbalanced traffic}
	\label{fig:th_shuffle_unbalanced_traffic}
	\endminipage
\end{figure*}

Simulations are done for a ($64\times 64$) switch with full depth ($M=k$) and a variable speedup factor. In this sub-section, we use the terminology Dynamic and Static to present the Clos-UDN performance with respectively a dynamic RR dispatching scheme and a static packets dispatching process. There are many ways to configure the first two stages connections. We choose to keep the default Clos interconnection as Fig.\ref{fig:modulo_shuffle} depicts. Considering a directional traffic, the $modulo XY$ algorithm would work as following: All packets that come from an input FIFO($i, r$) heading to OP($i,h$) are always forwarded to the ingress $i$ of CM($r$) via LI($i, r$). Resolving the packets destination would result in a direct route across the UDN fabric (the route from an input to an output of the NoC mesh with no turns). In case of crossing traffic, where packets stored in FIFO($i,r$) are destined to OP($j,h$) ($j\neq i$), the load is distributed in the UDN fabric and forwarded using different paths to the right LC($r, j$).
\vspace{0.2cm}


\begin{figure*}[tbh] 
	\begin{center}
		\includegraphics[width=4.7in]{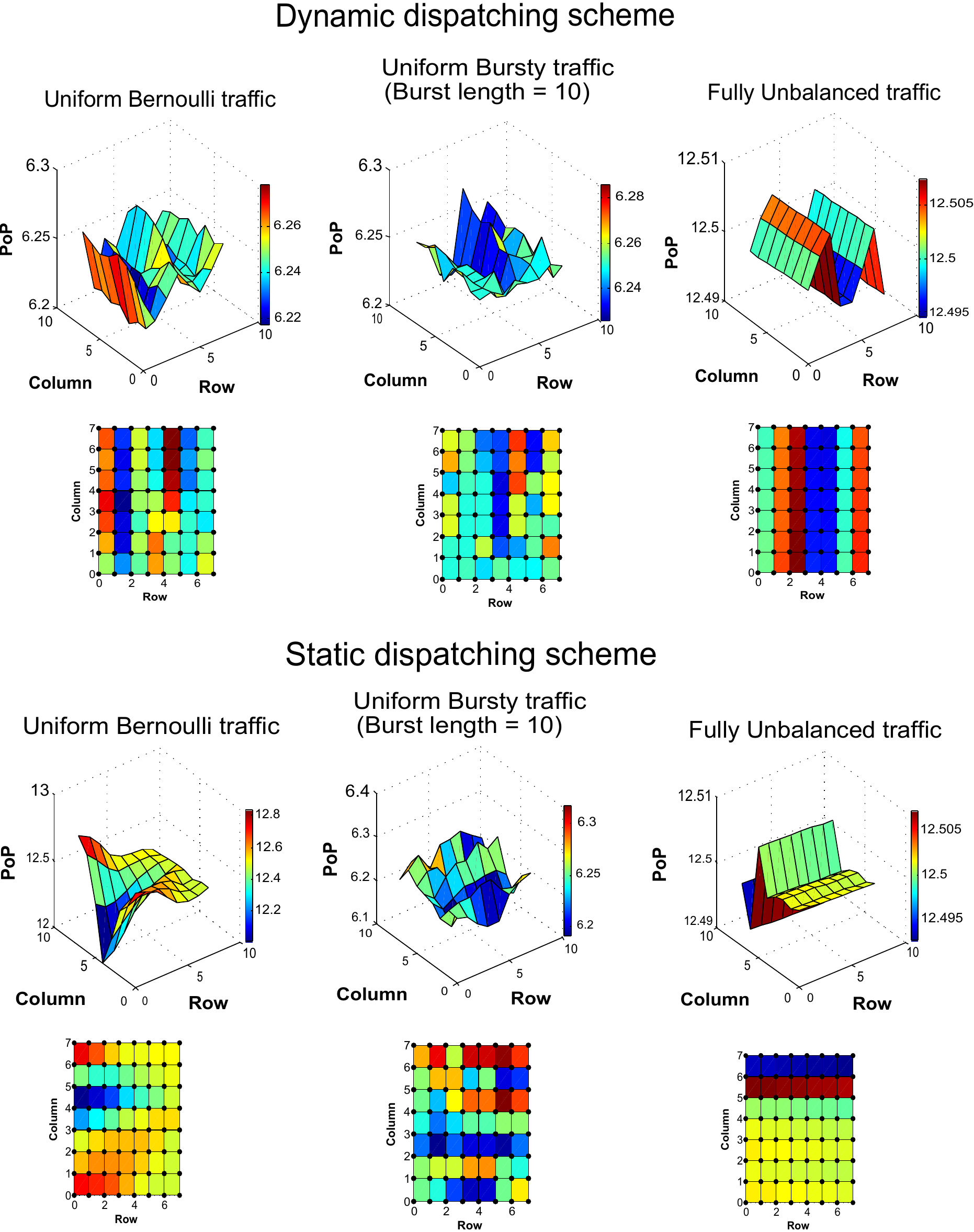}
		\vspace{.2cm}
		\caption{Propotion of Packets moving west-east in the middle-stage UDNs}
		\label{fig:link_east_static_dynamic}
	\end{center}
\end{figure*}

\paragraph{Uniform traffic}
Under uniform traffic, the two-stage Clos-UDN switch gives good throughput and bearable average packets delays. Fig.~\ref{fig:shuffle_uniform_traffic} compares the delay performance of the three-stage Clos-UDN and the two-stage Clos-UDN with in-order packets delivery guarantee. By adopting a static dispatching scheme, we reduce the number of simultaneously available links between any pair of IM/OM. This accounts for a throughput degradation that we clearly notice when the central switching modules are run at $SP=1$. Speeding up the CMs preserves the switch throughput and $SP\geq2$ makes the two-stage Clos-UDN achieve performance comparable to that of a three-stage Clos-UDN switch using a RR packets dispatching. 

Likewise, we simulated the two-stage switch with different $SP$ values under the uniform bursty traffic. The bursty traffic can be modulated as an an on-off Markov process, where the average burst length is set to 10. Fig.~\ref{fig:shuffle_bursty_traffic}  shows that the Clos-UDN with with static configuration and $SP=2$ provides 50$\%$ throughput. Setting $SP$ to 4 makes the switch reach a delay performance little higher than the switch with a dynamic RR dispatching.


\vspace{0.2cm}
\paragraph{Unbalanced traffic}
The unbalanced traffic is defined using an unbalanced coefficient $\omega$ that reflects the proportionality of the traffic distribution among the outputs. For an ($N\times N$) switch, we define the traffic load from an input port $s$ to an output port $d$ by $\rho_{s,d}$, where
\[ \rho_{s,d} =
\begin{cases}
\rho(\omega + \frac{1- \omega}{N})  & \quad \text{if } s = d \\
\rho\frac{1- \omega}{N}  & \quad \text{ otherwise}\\
\end{cases}
\]

Fig.~\ref{fig:shuffle_unbalanced_traffic} shows that using a speedup of one, the two-stage switch reaches up to $87\%$ throughput. Making $SP\geq2$ proves to be sufficient in the sens that it makes the statically configured switch achieve full throughput. We note that for high traffic loads, the average cell delay of the two-stage switch becomes better than the three-stage Clos-UDN with dynamic packets dispatching as depicted in Fig.~\ref{fig:zoom_unbal}.

Actually, the dynamic cells dispatching constantly disburses the traffic through LI links which results in good load balancing within the different UDN modules. On the contrary, a static scheme makes the load partition between CMs strongly dependent on the traffic type. If the switch is fed with skewed traffic, some LI links (and consequently UDNs) might be loaded. We intercept packets exiting the Clos-UDN switch and analyze the Proportion of Packets (PoP) going over East links of the  of the UDN' mini-routers. Fig.~\ref{fig:link_east_static_dynamic} illustrates the average PoP of the West-East traffic calculated over $8$ UDNs (in a ($64 \times64$) Clos-UDN switch operating with UDN $SP=2$). We note that packets get equally distributed among East links for both dispatching schemes under uniform traffic arrivals and that for a critical diagonal traffic (where input $i$ of the switch sends traffic only to output of the same index), the dynamic RR dispatching contributes to better load distribution in the UDN modules.  
\vspace{0.2cm}
\paragraph{Throughput stability in the two-stage Clos-UDN}
Limiting the speedup factor to $SP = 1$, limits the switch performance under a non-uniform traffic using both static and dynamic dispatching methods. The UDN fabrics being tanked with arriving packets cannot afford full throughput with a $SP=1$. Slightly increasing the UDNs speedup enhances the switch performance and $SP=2$ makes the two-stage Clos-UDN reach $100\%$ throughput under the complete range of $\omega$ as Fig.~\ref{fig:th_shuffle_unbalanced_traffic} depicts.

\section{Conclusion}\label{Sec.Conc}
In this paper, we propose a novel multistage switching architecture for Data Center Networks. The Clos-UDN is a highly-scalable and easily configurable switch with simple FIFO queuing at the input modules and simple packets dispatching schemes. We plug NoC-based fabric modules with on-chip buffering and arbitration in the middle stage of the Clos-network. The NoC switches, allow a pipelined and distributed scheduling and obviate the need for a centralized and complex arbiter as it is the case for bufferless and semi-buffered architectures (Concurrent dispatching for S$^3$ switch and CRRD matching for MSM switch). Our design also avoids the use of large crosspoint buffers like those that fully buffered structures require (MMM switch). 

We present and discuss the performance of two possible packets dispatching schemes. The Clos-UDN switch with a dynamic dispatching process mis-sequences packets delivery in the same way MMM does. We observe that it is possible to prevent packets from getting dis-ordered at first place by introducing a static configuration of the IM and CM modules interconnections. Although this reduces the switch architecture to two-stage Clos-network, this approach results in constantly dispatching packets of a given flow to the same CMs where they get forwarded in-order and in a multi-hop way until their output ports using deterministic routes.

Our extensive and detailed simulations show that the three-stage Clos-UDN provides high and stable throughput. Considering a minimum $SP=2$, the proposed switch gives good average latency as compared to MSM and MMM switches under different traffic loads with far less complex architecture and scheduling process. The Clos-UDN demonstrates: 1) a robustness of packet delay to the switch valency; 2) an immunity of overall delay in presence of hotspots; 3) almost constant delay variation under medium-to-high loads no matter the switch size and the traffic type are and 4) high achievable throughput. Based on our previous works and the current technology advances, we conjecture that running the switch central modules (UDNs) with speedup of two is quite straightforward. The HW implementation and prototyping of the Clos-UDN switch are reserved for future work.

\section{Acknowledgment}\label{Sec.Ack}
This work was supported by the EU Marie Curie Grant
(SCALE: PCIG-GA-2012-322250).  

\ifCLASSOPTIONcaptionsoff
  \newpage
\fi

\nocite{*}
\bibliographystyle{IEEEtran}
\bibliography{references}

\begin{thebibliography}{10}
\providecommand{\url}[1]{#1}
\csname url@samestyle\endcsname
\providecommand{\newblock}{\relax}
\providecommand{\bibinfo}[2]{#2}
\providecommand{\BIBentrySTDinterwordspacing}{\spaceskip=0pt\relax}
\providecommand{\BIBentryALTinterwordstretchfactor}{4}
\providecommand{\BIBentryALTinterwordspacing}{\spaceskip=\fontdimen2\font plus
\BIBentryALTinterwordstretchfactor\fontdimen3\font minus
  \fontdimen4\font\relax}
\providecommand{\BIBforeignlanguage}[2]{{%
\expandafter\ifx\csname l@#1\endcsname\relax
\typeout{** WARNING: IEEEtran.bst: No hyphenation pattern has been}%
\typeout{** loaded for the language `#1'. Using the pattern for}%
\typeout{** the default language instead.}%
\else
\language=\csname l@#1\endcsname
\fi
#2}}
\providecommand{\BIBdecl}{\relax}
\BIBdecl

\bibitem{art51}
\BIBentryALTinterwordspacing
``Cisco,'' 2016. [Online]. Available:
  \url{http://www.cisco.com/c/en/us/products/switches/nexus-5000-series-switches/datasheet-listing.html}
\BIBentrySTDinterwordspacing

\bibitem{art1}
N.~I. Chrysos, ``Request-{G}rant {S}cheduling for {C}ongestion {E}limination in
  {M}ulti-{S}tage {N}etworks,'' Crete University, 2006, Tech. Rep.

\bibitem{art17}
N.~Chrysos and M.~Katevenis, ``Scheduling in {N}on-{B}locking {B}uffered
  {T}hree-stage {S}witching {F}abrics.'' in \emph{INFOCOM on}, vol.~6, 2006,
  pp. 1--13.

\bibitem{art18}
H.~J.~C. Yu~Xia, ``On {P}ractical {S}table {P}acket {S}cheduling for
  {B}ufferless {T}hree-stage {C}los-network switches,'' in \emph{HPSR.}\hskip
  1em plus 0.5em minus 0.4em\relax IEEE, 2013, pp. 7--14.

\bibitem{art50}
\BIBentryALTinterwordspacing
``Dell,'' May 2016. [Online]. Available:
  \url{http://i.dell.com/sites/doccontent/shared-content/data-sheets/en/Documents/Dell_Networking_S4048T_ON_Spec_Sheet.pdf}
\BIBentrySTDinterwordspacing

\bibitem{art52}
\BIBentryALTinterwordspacing
``Juniper {N}etworks,'' June 2015. [Online]. Available:
  \url{http://www.juniper.net/assets/us/en/local/pdf/datasheets/1000414-en.pdf}
\BIBentrySTDinterwordspacing

\bibitem{art2}
C.~Clos, ``A {S}tudy of {N}on-{B}locking {S}witching {N}etworks,'' \emph{Bell
  System Technical Journal on}, vol.~32, no.~2, pp. 406--424, 1953.

\bibitem{art8}
H.~J. Chao, J.~Park, S.~Artan, S.~Jiang, and G.~Zhang, ``True{W}ay: {A}
  {H}ighly {S}calable {M}ulti-plane {M}ulti-stage {B}uffered packet switch,''
  in \emph{HPSR.}\hskip 1em plus 0.5em minus 0.4em\relax IEEE, 2005, pp.
  246--253.

\bibitem{art16}
E.~Oki, N.~Kitsuwan, and R.~Rojas-Cessa, ``Analysis of {S}pace-{S}pace-{S}pace
  {C}los-network packet switch,'' in \emph{ICCCN.}\hskip 1em plus 0.5em minus
  0.4em\relax IEEE, 2009, pp. 1--6.

\bibitem{art7}
Z.~Dong, R.~Rojas-Cessa, and E.~Oki, ``Memory-{M}emory-{M}emory {C}los-network
  packet switches with {I}n-sequence {S}ervice,'' in \emph{HPSR.}\hskip 1em
  plus 0.5em minus 0.4em\relax IEEE, 2011, pp. 121--125.

\bibitem{art3}
F.~M. Chiussi, J.~G. Kneuer, and V.~P. Kumar, ``Low-cost {S}calable switching
  solutions for broadband networking: the {ATLANTA} architecture and
  {C}hipset,'' \emph{Communications Magazine on}, vol.~35, no.~12, pp. 44--53,
  1997.

\bibitem{art6}
X.~Li, Z.~Zhou, and M.~Hamdi, ``Space-{M}emory-{M}emory architecture for
  {C}los-network packet switches,'' in \emph{ICC.}\hskip 1em plus 0.5em minus
  0.4em\relax IEEE, 2005, pp. 1031--1035.

\bibitem{art4}
E.~Oki, Z.~Jing, R.~Rojas-Cessa, and H.~J. Chao, ``Concurrent
  {R}ound-{R}obin-based {D}ispatching schemes for {C}los-network switches,''
  \emph{IEEE/ACM on}, vol.~10, no.~6, pp. 830--844, 2002.

\bibitem{art5}
J.~Kleban and A.~Wieczorek, ``{CRRD-OG}: {A} packet {D}ispatching algorithm
  with {O}pen {G}rants for {T}hree-stage {B}uffered {C}los-network switches,''
  in \emph{HPSR.}\hskip 1em plus 0.5em minus 0.4em\relax IEEE, 2006, pp. 6--pp.

\bibitem{art13}
K.~Goossens, L.~Mhamdi, and I.~V. Senin, ``Internet-router {B}uffered
  {C}rossbars based on {N}etworks on {C}hip,'' in \emph{DSD.}\hskip 1em plus
  0.5em minus 0.4em\relax IEEE, 2009, pp. 365--374.

\bibitem{art23}
E.~Bastos, E.~Carara, D.~Pigatto, N.~Calazans, and F.~Moraes, ``{MOTIM-A}
  {S}calable {A}rchitecture for {E}thernet switches,'' in \emph{ISVLSI.}\hskip
  1em plus 0.5em minus 0.4em\relax IEEE, 2007, pp. 451--452.

\bibitem{art24}
F.~Moraes, N.~Calazans, A.~Mello, L.~M{\"o}ller, and L.~Ost, ``{HERMES}: {A}n
  {I}nfrastructure for {L}ow {A}rea overhead packet-switching {N}etworks on
  {C}hip,'' \emph{INTEGRATION, the VLSI journal on}, vol.~38, no.~1, pp.
  69--93, 2004.

\bibitem{art15}
T.~Karadeniz, L.~Mhamdi, K.~Goossens, and J.~Garcia-Luna-Aceves, ``Hardware
  {D}esign and {I}mplementation of a {N}etwork-on-{C}hip based load balancing
  switch fabric.'' in \emph{ReConFig.}, 2012, pp. 1--7.

\bibitem{art14}
L.~Mhamdi, K.~Goossens, and I.~V. Senin, ``Buffered {C}rossbar {F}abrics
  {B}ased on {N}etworks on {C}hip.'' in \emph{CNSR.}, 2010, pp. 74--79.

\bibitem{art22}
A.~Bitar, J.~Cassidy, N.~E. Jerger, and V.~Betz, ``Efficient and {P}rogrammable
  {E}thernet switching with a {N}o{C}-enhanced {FPGA},'' in \emph{Proceedings
  of the 10th ACM/IEEE ANCS}.\hskip 1em plus 0.5em minus 0.4em\relax ACM, 2014,
  pp. 89--100.

\bibitem{art36}
F.~Hassen and L.~Mhamdi, ``A {M}ulti-{S}tage {P}acket-{S}witch {B}ased on
  {N}o{C} {F}abrics for {D}ata {C}enter {N}etworks,'' in \emph{Globecom
  Workshops}.\hskip 1em plus 0.5em minus 0.4em\relax IEEE, 2015, pp. 1--6.

\bibitem{art32}
I.~Keslassy and N.~McKeown, ``Maintaining {P}acket {O}rder in {T}wo-{S}tage
  switches,'' in \emph{INFOCOM. 21st Annual Joint Conference of the IEEE
  Computer and Communications Societies on}, vol.~2.\hskip 1em plus 0.5em minus
  0.4em\relax IEEE, 2002, pp. 1032--1041.

\bibitem{art34}
Z.~Dong, R.~Rojas-Cessa, and E.~Oki, ``Buffered {C}los-network {P}acket
  {S}witch with per-output flow queues,'' \emph{Electronics letters on},
  vol.~47, no.~1, pp. 32--34, 2011.

\bibitem{art31}
C.-S. Chang, D.-S. Lee, and Y.-S. Jou, ``Load balanced {B}irkhoff-{V}on
  {N}eumann switches,'' in \emph{HPSR Workshop}.\hskip 1em plus 0.5em minus
  0.4em\relax IEEE, 2001, pp. 276--280.

\bibitem{art33}
R.~Roberto and C.~Lin, ``Scalable {T}wo-stage {C}los-network {S}witch and
  {M}odule-{F}irst {M}atching,'' in \emph{HPSR Workshop.}, 2006, pp. 6--11.

\bibitem{art12}
K.~Goossens, J.~Dielissen, and A.~Radulescu, ``{\AE}thereal {N}etwork on
  {C}hip: {C}oncepts, {A}rchitectures, and {I}mplementations,'' \emph{Design \&
  Test of Computers on}, vol.~22, no.~5, pp. 414--421, 2005.

\bibitem{art53}
A.~Radulescu, J.~Dielissen, S.~G. Pestana, O.~P. Gangwal, E.~Rijpkema,
  P.~Wielage, and K.~Goossens, ``An {E}fficient {O}n-{C}hip {NI} {O}ffering
  {G}uaranteed {S}ervices, {S}hared-{M}emory {A}bstraction, and {F}lexible
  {N}etwork {C}onfiguration,'' \emph{IEEE Transactions on {C}omputer-{A}ided
  {D}esign of {I}ntegrated {C}ircuits and {S}ystems}, vol.~24, no.~1, pp.
  4--17, 2005.

\bibitem{art26}
Z.~Dong and R.~Rojas-Cessa, ``Non-blocking {M}emory-{M}emory-{M}emory
  {C}los-network packet switch,'' in \emph{Sarnoff Symposium, 34th IEEE}.\hskip
  1em plus 0.5em minus 0.4em\relax IEEE, 2011, pp. 1--5.

\bibitem{art27}
S.-T. Chuang, S.~Iyer, and N.~McKeown, ``Practical {A}lgorithms for
  {P}erformance {G}uarantees in {B}uffered {C}rossbars,'' in \emph{INFOCOM.
  24th Annual Joint Conference of the IEEE Computer and Communications
  Societies on}, vol.~2.\hskip 1em plus 0.5em minus 0.4em\relax IEEE, 2005, pp.
  981--991.

\bibitem{art29}
W.~Song, D.~Edwards, J.~Garside, and W.~J. Bainbridge, ``Area {E}fficient
  {A}synchronous {SDM} routers using 2-stage {C}los switches,'' in
  \emph{Proceedings of the Conference on Design, Automation and Test in
  Europe}.\hskip 1em plus 0.5em minus 0.4em\relax EDA Consortium, 2012, pp.
  1495--1500.

\bibitem{art35}
C.-S. Chang, D.-S. Lee, and Y.-S. Jou, ``Load balanced {B}irkhoff-{V}on
  {N}eumann switches, part i: one-stage buffering,'' \emph{Computer
  Communications on}, vol.~25, no.~6, pp. 611--622, 2002.

\bibitem{speedup}
S.-T. Chuang, A.~Goel, N.~McKeown, and B.~Prabhakar, ``Matching output queueing
  with a combined input/output-queued switch,'' \emph{Selected Areas in
  Communications, IEEE Journal on}, vol.~17, no.~6, pp. 1030--1039, 1999.

\end{thebibliography}

\newpage
\begin{IEEEbiography}[{\includegraphics[width=1in,height=1.25in,clip,keepaspectratio]{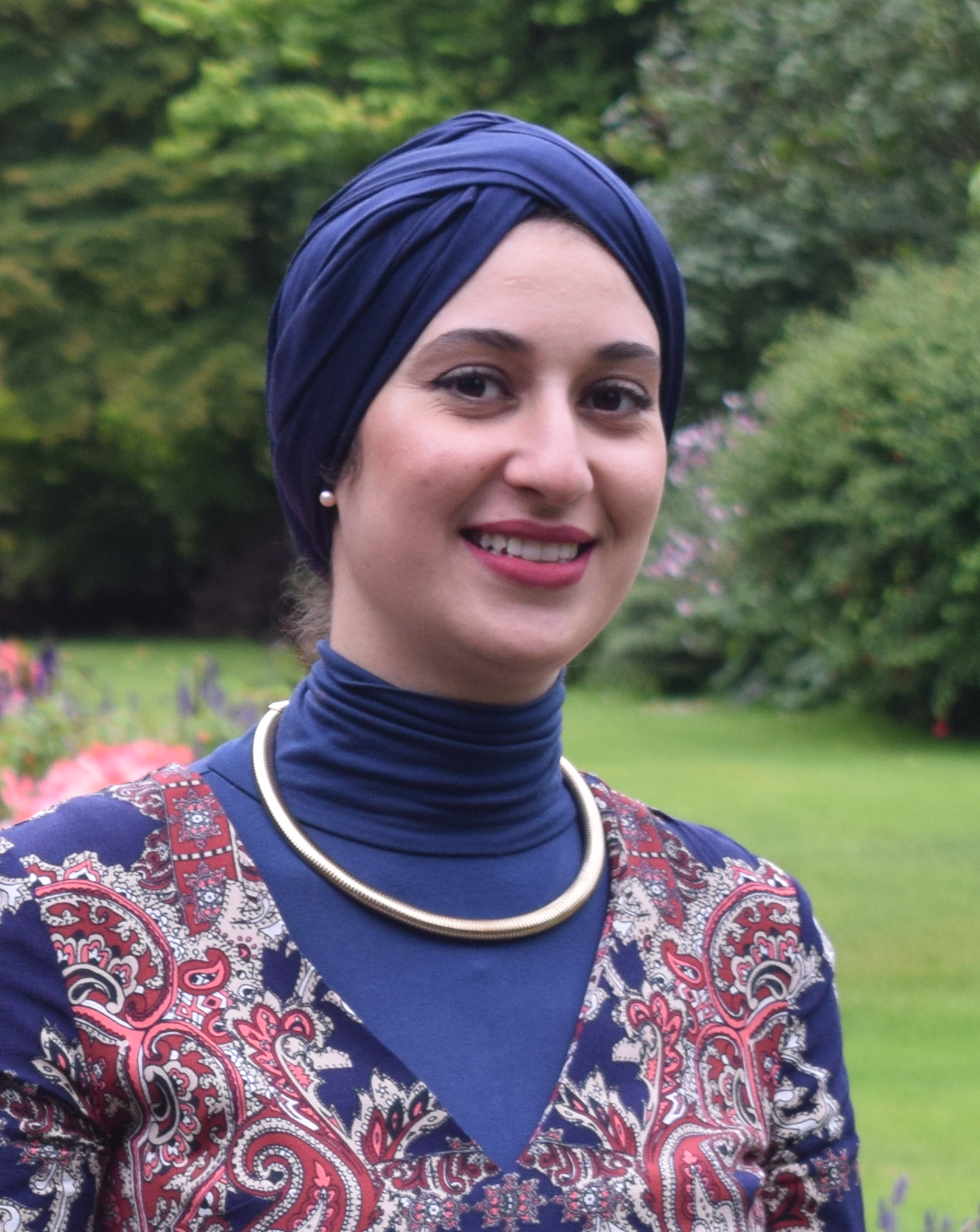}}]{Fadoua HASSEN}
received her M.S. degree in Telecommunications – Communication engineering (with distinction) from the Higher School of Communication of Tunis, SUP'COM – University of Carthage, in 2011. She is currently working towards the Ph.D degree in Electrical Engineering at the University of Leeds. Her research interests include high-performance packet-switch design, scalable switching architectures and switching/routing in Data Center Networks. She is a student member of the IEEE.
\end{IEEEbiography}


\begin{IEEEbiography}[{\includegraphics[width=1in,height=1.25in,clip,keepaspectratio]{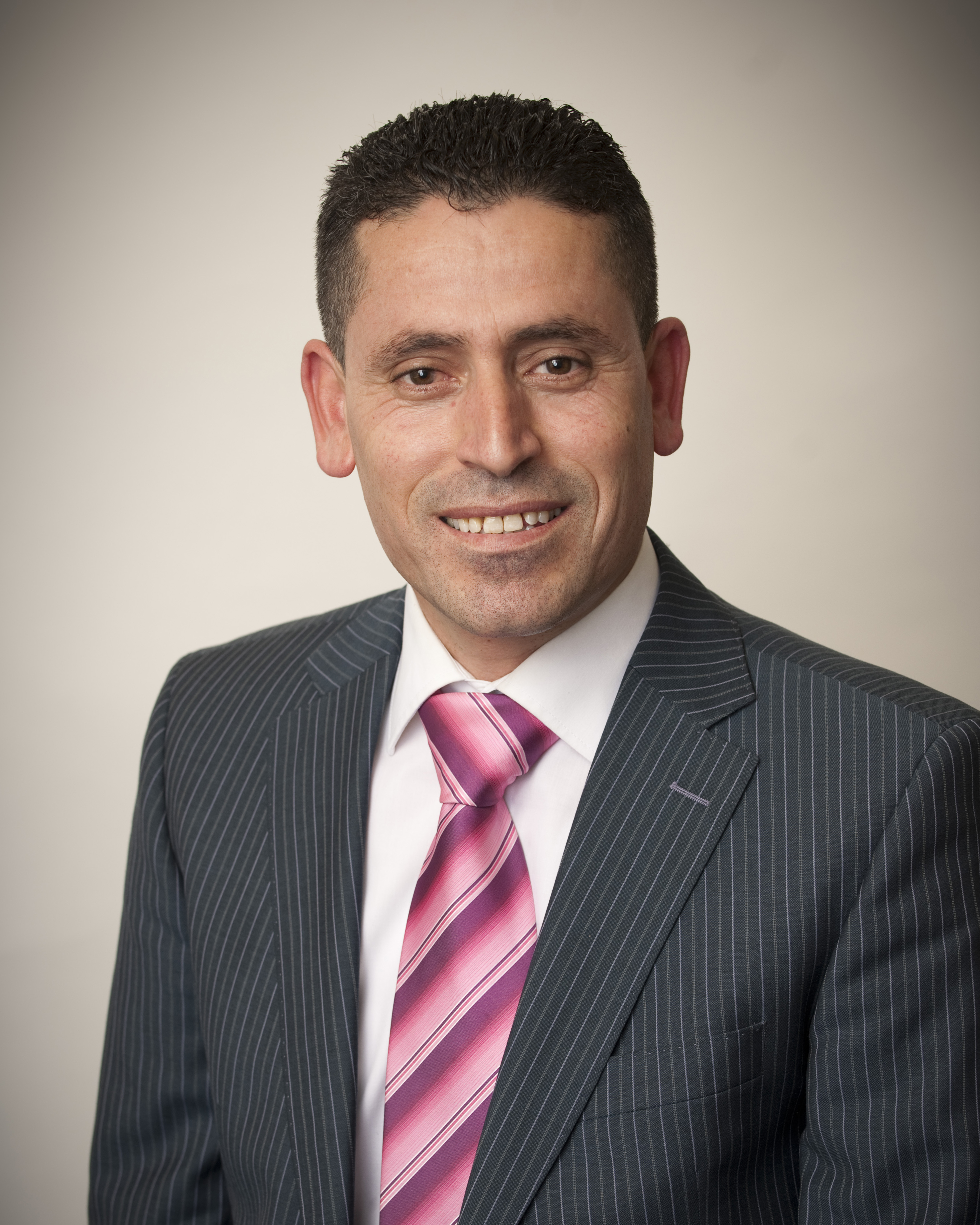}}]{Lotfi MHAMDI}
received the Master of Philosophy (MPhil.) degree in computer science from the Hong Kong University of Science and Technology (HKUST) in 2002 and the PhD. degree in computer engineering from Delft University of Technology (TU Delft), The Netherlands, in 2007. He continued his work at TU Delft as post-doctoral researcher, working on high-performance networking tropics within various European Union funded research projects. Since July 2011, he has been a Lecturer with the school of Electronic and Electrical Engineering at the University of Leeds, UK. Dr. Mhamdi is/was a technical program committee member in various conferences, including the IEEE International Conference on Communications (ICC), the IEEE GLOBECOM, the IEEE Workshop on High Performance Switching and Routing (HPSR), and the ACM/IEEE International Symposium on Networks-on-Chip (NoCS). His research work spans the area of high-performance networks including the architecture, design, analysis, scheduling, and management of high-performance switches and Internet routers. He is a member of the IEEE.
\end{IEEEbiography}

\end{document}